\documentclass[12pt]{iopart}
\usepackage{iopams}
\usepackage{graphicx,epsf}

\newcommand{\laeq}{\ \raisebox{-.7ex}{$\stackrel{\textstyle<}{\sim}$}\ }
\newcommand{\gaeq}{\ \raisebox{-.7ex}{$\stackrel{\textstyle>}{\sim}$}\ }
				   
\begin{document}

%\jl{14} %what journal is Nuc Fus?

\title{Electron fishbones: theory and experimental evidence}

\author{F. Zonca$^1$, P. Buratti$^1$, A. Cardinali$^1$, L. Chen$^{2,3}$, J.-Q. Dong$^4$, Y.-X. Long$^4$, A.V. Milovanov$^{1,5,6}$, F. Romanelli$^1$, P. Smeulders$^1$, L. Wang$^7$, Z.-T. Wang$^4$, C. Castaldo$^1$, R. Cesario$^1$, E. Giovannozzi$^1$, M. Marinucci$^1$ and V. Pericoli Ridolfini$^1$}

\address{~$^1$Associazione Euratom-ENEA sulla Fusione, C.P. 65 - I-00044 - Frascati, Italy}
\address{~$^2$Dept. of Physics and Astronomy,  Univ. of California, Irvine CA 92697-4575, U.S.A.}
\address{~$^3$Inst. for Fusion Theory and Simulation, Zhejiang Univ., Hangzhou 310027, P.R.C.}
\address{~$^4$Southwestern Institute of Physics, P.O. Box 432, Chengdu 610041, P.R.C.}
\address{~$^5$Dept. of Physics and Technology, Univ. of Troms\o, N-9037 Troms\o, Norway}
\address{~$^6$Space Research Institute, Russian Academy of Sciences, Moscow, Russia}
\address{~$^7$Institute of Physics, Chinese Academy of Sciences, Beijing 100080, P.R.C.}

\begin{abstract}
We discuss the processes underlying the excitation of fishbone-like internal kink instabilities driven by supra-thermal electrons generated experimentally by  different means: Electron Cyclotron Resonance Heating (ECRH) and by Lower Hybrid (LH) power injection. The peculiarity and interest of exciting these electron fishbones by ECRH only or by LH only is also analyzed.  Not only the mode stability is explained, but also the transition between steady state nonlinear oscillations to bursting (almost regular) pulsations, as observed in FTU, is interpreted in terms of the LH power input. These results are directly relevant to the investigation of trapped alpha particle interactions with low-frequency MHD modes in burning plasmas: in fact, alpha particles in reactor relevant conditions are characterized by small dimensionless orbits, similarly to electrons; the trapped particle bounce averaged dynamics, meanwhile, depends on energy and not mass.
\end{abstract} 

%Magnetohydrodynamic waves (e.g., Alfven waves)
%Nonlinear phenomena: waves, wave propagation, and other interactions (including parametric effects, mode coupling, ponderomotive effects, etc.)
%Fusion products effects (e.g., alpha-particles, etc.), fast particle effects
%Ideal and resistive MHD modes; kinetic modes
\submitted
\maketitle

\section{Introduction and Background}
\label{sec:intro}

Fishbone - like internal kink instabilities driven by electrons have been observed for the first time on DIII-D in conjunction with Electron Cyclotron Resonance Heating (ECRH) on the high field side~\cite{wong00}.  There, the excitation was attributed to {\it barely trapped supra-thermal electrons}, which are characterized by {\it drift-reversal} and can destabilize a mode propagating in the ion diamagnetic direction in the presence of an {\it inverted spatial gradient} of the supra-thermal tail.
Similar but higher frequency modes were observed in Compass-D~\cite{valovic00} during ECRH and Lower Hybrid (LH) power injection, with {\it chirping} frequency comparable with that of the Toroidal Alfv\'{e}n Eigenmode~\cite{ccc85} (TAE),  $\omega \laeq \omega_{TAE}$. Observations of {\it electron fishbones} with ECRH only~\cite{ding02,li02} and LH only~\cite{smeulders02,romanelli02} have been also reported in  HL-1M and FTU, respectively. More recently, electron fishbones have been observed in Tore Supra~\cite{maget06} due to resonant excitation of a double-kink mode by supra-thermal electrons generated with LH power injection.

In the present work, we analyze the peculiar features of electron fishbones versus those of the well known ion fishbone~\cite{mcguire83,chen84,coppi86}. Due to the frequency gap in the low-frequency shear Alfv\'{e}n continuum for modes propagating in the ion diamagnetic direction~\cite{coppi86}, effective electron fishbone excitation favors conditions characterized by supra-thermal electron drift reversal, consistently with experimental observations. For the same reason, the spatial gradient inversion of the supra-thermal electron tail is necessary, explaining why ECRH excitation is observed with high field side deposition only~\cite{wong00,ding02,li02,zhou05,wang06}. Here, we also discuss the peculiar roles of {\it circulating supra-thermal electrons} for electron fishbone excitations with LH only: the {\it barely circulating} population providing directly the mode drive and the {\it well circulating} particles controlling both the drift-reversal condition as well as the ideal MHD stability via their effect on the plasma current profile. The role of LH current drive in controlling sawtooth oscillations via the local magnetic shear at the $q=1$ surface ($q$ being the safety factor) was recently documented by the HT-7 tokamak~\cite{sun05b}.

As in the case of  ion fishbones, two branches of the electron fishbone are shown to exist: a discrete gap mode~\cite{coppi86} and a continuum resonant mode~\cite{chen84}. Contrary to the gap mode, the continuum resonant mode can propagate in the electron diamagnetic direction as well. Thus, it does not require neither drift-reversal nor inverted spatial gradient of the supra-thermal electron tail. However, its threshold condition in this case is higher and it requires high power densities to be excited. So, even the case of the continuum resonant fishbone mode tends to favor the branch propagating in the ion diamagnetic direction, which minimizes continuum damping. If the effective temperature of the supra-thermal electron tail is sufficiently high, the present theory predicts that fishbone oscillations can be excited at frequencies comparable with those typical of the Geodesic Acoustic Mode (GAM)~\cite{winsor68} or the Beta induced Alfv\'{e}n Eigenmode (BAE)~\cite{heidbrink93,turnbull93}. Unlike the case of fishbone gap modes in the ion diamagnetic gap~\cite{coppi86} of the low-frequency shear Alfv\'{e}n continuum, fishbone gap modes in the BAE gap~\cite{chu92} do not favor the propagation in the ion diamagnetic direction, since the gap structure is nearly symmetric in frequency~\cite{zonca96}.
Here, we discuss these issues using one single general {\it fishbone-like} dispersion relation~\cite{zonca06,chen06}, describing mode excitation by trapped as well as circulating supra-thermal electrons in both monotonic and reversed magnetic shear equilibria~\cite{hastie87}.

In this work, we also analyze the nonlinear physics of electron fishbones, of which  FTU experimental results provide a nice and clear example (see Figure~1): during high power LH injection, an evident transition in the  electron fishbone signature takes place from almost steady state nonlinear oscillations (fixed point) to regular bursting behavior (limit cycle). Here, we present a simple yet relevant nonlinear dynamic model for predicting and interpreting these observations.
\begin{figure}
\begin{center}
\noindent\epsfxsize=0.75\linewidth\epsfbox{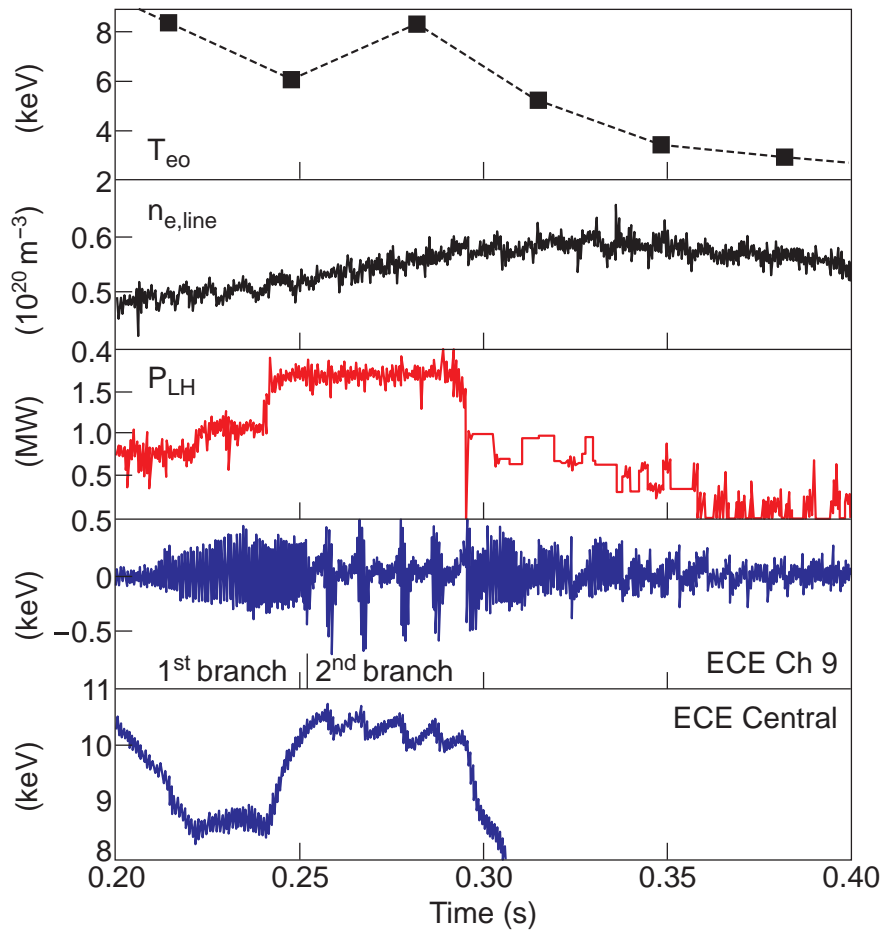}
\end{center}
\label{fig:figure1}
\caption{Time evolution (from top to
bottom) of thermal electron temperature,
plasma line density, LH coupled power,
fast electron temperature fluctuations and central
radiation temperature in FTU shot \# 20865. It is clear that the nonlinear behavior of fast electron temperature fluctuations (electron fishbone) reflects the level of LH power input.}  
\end{figure}

These results are directly relevant to the investigation of trapped alpha particle
interactions with low-frequency MHD modes in burning plasmas: in fact, alpha
particles in reactor relevant conditions are characterized by small dimensionless orbits,
similarly to electrons; the trapped particle bounce averaged dynamics, meanwhile,
depends on energy and not mass. Rigorously speaking, the same argument applies to {\em barely circulating} particles as well, whose definition is given in Section~\ref{sec:disprel}. For these reasons, we could draw a symmetry between trapped ion (alpha particle) and trapped electron dynamics and exploit the combined experimental use of LH and ECRH analogously to what is done with Neutral Beam Injection and Ion Cyclotron Resonance Heating.

\section{Mode dispersion relations}
\label{sec:disprel}

The fishbone dispersion relation can be obtained by the standard matching procedure of mode structures in the ideal region and inertial layer~\cite{coppi66} and generalizing the results therein. Here, we choose to solve quasi-neutrality and vorticity equations following the procedure of Ref.~\cite{pegoraro86}, where the solution of the kinetic layer equations in the Fourier space are matched to the ideal region. Letting $x=-k_\theta (r-r_s)$, with $k_\theta$ the poloidal wave vector (here, $k_\theta r_s=-1$) and $r_s$ the radius of the $q(r_s)=1$ surface, we introduce the representation
 \begin{equation}
 \delta \phi (x) = \int d\eta e^{- {\rm i} \eta x} \delta \Phi(\eta) \;\; \label{eq:fourier}
 \end{equation}
for the scalar potential fluctuation and other fields. For finite shear, $s=r_s q_s^{\prime}/q_s$, the asymptotic ideal region solution for odd parity modes when approaching the inertial layer is~\cite{pegoraro86} 
\begin{equation}
|\eta| \delta \Phi \simeq {- \rm i} (\Delta \delta \phi/2\pi) (k_\theta/|k_\theta|) \left(\eta/|\eta|\right)  \left( 1 + |\eta| \delta \hat {W} / s^2 \right) \;\; , \label{eq:ideal0}
\end{equation}
for $1\ll |\eta| \ll |\gamma/\omega|^{-1}$, with $\gamma/\omega$ the normalized mode growth rate and $\delta \hat {W}$ the normalized potential energy, which, for low-pressure and high aspect ratio tokamak equilibria with circular flux surfaces, is given by $\delta \hat {W}=(2 |k_\theta| R_0/B_0^2)(\delta W/|\Delta \delta \xi_r|^2) (q^2(r_s)/r_s)$~\cite{chen84}. Here, $R_0$ is the tokamak major radius, $B_0$ the on axis magnetic field, $\Delta \delta \xi_r$ is the jump of the radial mode displacement across the inertial layer (with $\Delta \delta \phi$ the corresponding jump in $\delta \phi$) and $\delta W$ the mode potential energy. Meanwhile, the inertial (kinetic) layer solution is~\cite{zonca06} 
\begin{equation}
|\eta| \delta \Phi \simeq  {- \rm i} (\Delta \delta \phi/2\pi) (k_\theta/|k_\theta|) \left(\eta/|\eta|\right) \left( 1 + {\rm i} |\eta| \Lambda/ |s| \right) \;\; . \label{eq:kine0}
\end{equation}
Here, $\Lambda$ is the generalized inertia term introduced in~\cite{zonca06}. Given Eqs.~(\ref{eq:ideal0}) and~(\ref{eq:kine0}), the mode dispersion relation reads~\cite{chen84,coppi86}
\begin{equation}
{\rm i}  \Lambda |s| = \delta \hat {W}  = \delta \hat {W}_f + \delta \hat {W}_k \;\; , \label{eq:fish0}
\end{equation}
where the fluid $\delta \hat {W}_f$, in its simplest expression, is given by~\cite{bussac75}
\begin{equation}
\delta \hat {W}_f = 3 \pi \Delta q_0 \left( 13/144 - \beta_{ps}^2 \right) \left(r_s^2/R_0^2\right)
\end{equation}
with $\beta_{ps}=-(R_0/r_s^2)^2 \int_0^{r_s} r^2 (d\beta/dr) dr$,  $\Delta q_0 = 1 - q(r=0)$ and $\beta = 8\pi P/B_0^2$ the ratio of kinetic and magnetic pressures. The fluid term, $\delta \hat{W}_f$, includes the contribution of the energetic (hot) particle adiabatic and convective responses as well~\cite{chen84}. Meanwhile, the kinetic $\delta \hat {W}_k$ is~\cite{chen84}
\begin{equation}
\hspace*{-2.4cm}\delta \hat{W}_k= 4 \frac{\pi^2}{B_0^2} m\omega_c^2 \frac{R_0}{r_s^2} \int_0^{r_s} \frac{r^3}{q} dr \int {\cal E} d {\cal E} d\lambda \sum_{v_\parallel/|v_\parallel| = \pm 1}\overline{e^{{\rm i} q(r)\theta}\omega_d e^{-{\rm i} \theta}} \; \overline{e^{{\rm i} \theta }\omega_d e^{-{\rm i} q(r)\theta }} \frac{ \tau_b \, Q F_0}{\bar{\omega}_d - \omega} \;\; , \label{eq:deltawk}
\end{equation}
where $m$ is the energetic particle mass, $\omega_c = (eB/mc)$ is the cyclotron frequency, ${\cal E}=v^2/2$, $\lambda=\mu B_0/{\cal E}=(B_0/B) v_\bot^2/v^2$, ${\bf B}\cdot {\bf \nabla} (\zeta - q(r) \theta)=0$, $\zeta$ is the ``toroidal angle'' chosen such that $(r,\theta,\zeta)$ is a toroidal flux coordinate system with straight field lines ($q=q(r)$), 
$\overline{(...)} = (\oint v_\parallel^{-1} d\ell)^{-1} \oint v_\parallel^{-1} (...) d\ell$ denotes bounce-averaging, $\ell$ is the arc length along the equilibrium ${\bf B}$-field, $\tau_b$ is the bounce/transit time for magnetically trapped/circulating particles, $\omega_d$ is the magnetic drift frequency and $QF_{0} = (\omega \partial_{\cal E} + \hat{\omega}_{*}) F_{0}$,  
$\hat{\omega}_{*} F_{0} = \omega_{c }^{-1}$(${\bf k} \times {\bf B}/B$)
$\cdot {\bf \nabla} F_{0}$, with $F_0=F_0({\cal E}, \lambda, v_\parallel/|v_\parallel)$ the fast particle equilibrium distribution function. 
In deriving Eq.~(\ref{eq:deltawk}), we have closely followed \cite{chen84} and solved for the energetic particle distribution function
\begin{equation}
\delta f = \frac{e}{m} \frac{\partial F_0}{\partial {\cal E}} \delta \phi + \delta H = \frac{e}{m} \frac{\partial F_0}{\partial {\cal E}} \delta \phi -  \frac{e}{m} \frac{Q F_0}{\omega} \delta \phi + \delta K \;\; , \label{eq:deltaf}
\end{equation}
neglecting finite orbit widths and separating both adiabatic ($\propto \partial_{\cal E} F_0$) as well as convective ($\propto Q F_0$) responses. In this way, one obtains
\begin{equation}
\delta K = \frac{e}{m} \frac{Q F_0}{\omega} \frac{\overline{e^{{\rm i} q(r)\theta}\omega_d e^{-{\rm i} \theta}}}{\bar{\omega}_d - \omega} \delta \phi_0 (r) e^{{\rm i} (\zeta - q(r)\theta )} \;\; , \label{eq:deltak}
\end{equation}
with $\delta \phi = \delta \phi_0 (r) \exp {\rm i} (\zeta - \theta)$. 
In this form, the dispersion relation neglects the thermal ion kinetic response in the ideal region~\cite{graves00}, whose analysis is outside the scope of this work and, for our purposes, we can consider as included in the expression of $\delta \hat {W}_f$ along with the contribution of the energetic (hot) particle adiabatic and convective responses~\cite{chen84}. Note that Eq.~(\ref{eq:deltawk}) depends only on the fast particle energy: the only residual mass dependence would be through their finite orbit width, which are neglected in the present treatment since we are assuming that the characteristic orbit size is much smaller than the fluctuation wavelength in the ideal region. This fact confirms our conjecture that experimental studies of electron fishbones are relevant for understanding alpha particle dynamics in burning plasmas, as noted in Section~\ref{sec:intro}. More detailed discussions on this issue are presented in Section~\ref{sec:NL}.

For the $s=0$ case but finite $S^2 \equiv r_s^2 q_s^{\prime \prime} / q_s^2$, Eq.~(\ref{eq:fish0}) cannot be applied; meanwhile, the asymptotic expressions corresponding to Eqs.~(\ref{eq:ideal0}) and~(\ref{eq:kine0}) are changed to
\begin{equation}
|\eta| \delta \Phi \simeq {- \rm i} (\Delta \delta \phi/2\pi) (k_\theta/|k_\theta|) \left(\eta/|\eta|\right)  \left( 1 - (2 k_\theta^2 r_s^2/3 q_s^2) |\eta|^3 \delta \hat {W} / S^4 \right) \; , \label{eq:ideal1}
\end{equation}
for the ideal region,  while the inertial (kinetic) layer solution becomes~\cite{zonca_easter03}
\begin{equation}
|\eta| \delta \Phi \simeq  {- \rm i} (\Delta \delta \phi/2\pi) (k_\theta/|k_\theta|) \left(\eta/|\eta|\right) \left( 1 + {\rm i} \alpha_1 \alpha_2 (\alpha_1+\alpha_2)|\eta|^3/6 \right) \;\; , \label{eq:kine1}
\end{equation}
where, $\alpha_1$ and $\alpha_2$ are defined as
\begin{eqnarray}
\alpha_1 & = & - \left( -\frac{2 |k_\theta r_s|}{q_s S^2} \left( \Lambda + k_{\| s} q_s R_0 \right) \right)^{1/2} \;\; , \nonumber \\
\alpha_2 & = &  \left( \frac{2 |k_\theta r_s|}{q_s S^2} \left( \Lambda - k_{\| s} q_s R_0 \right) \right)^{1/2} \;\; . \label{eq:alpha12}
\end{eqnarray}
By asymptotic matching procedure between Eqs.~(\ref{eq:ideal1}) and~(\ref{eq:kine1}), we readily derive the mode dispersion relation with a simple inertial layer at $r_s$~\cite{hastie87}
\begin{equation}
 - S \left( \Delta q_s^2 - \Lambda^2\right)^{3/4} \left[ 1 + \Delta q_s/ \sqrt{ \Delta q_s^2 - \Lambda^2 } \right]^{1/2} = \delta \hat {W}_f + \delta \hat {W}_k \;\;  ,  \label{eq:fish1}
\end{equation}
with $k_{\| s} q_s R_0 = \Delta q_s = q_s -1$ in this case.

Equations~(\ref{eq:fish0}) and~(\ref{eq:fish1}) are the basis for our linear stability studies of electron-fishbones. Their general structure in is known; however, we want to emphasize two novel aspects: (i) that Eq.~(\ref{eq:deltawk}) describes the resonant excitation of internal kink fluctuations by both trapped as well as {\em barely circulating} supra-thermal electron tails; (ii) that the analysis of the generalized inertia term, $\Lambda$~\cite{zonca06,chen06}, demonstrates the existence of ion- and electron-fishbones at frequencies comparable with that of GAM~\cite{winsor68} and BAE~\cite{heidbrink93,turnbull93}. That {\em well circulating} supra-thermal electron tails can control the internal kink stability via their influence on the $q$-profile, i.e. $\delta \hat{W}_f$, has been noted for explaining recent observations on the HT-7 tokamak~\cite{sun05b} and will be simply assumed in this work.

\subsection{Resonant excitation by trapped and barely circulating supra-thermal tails}
\label{sec:trapcirc}

For analyzing the different roles of trapped and circulating particles, we move from $({\cal E}, \lambda)$ to $({\cal E}, \kappa^2)$ space, with 
\begin{equation}
\kappa^2 = \frac{2(r/R_0) \lambda}{1 - (1-r/R_0) \lambda} \;\; , \label{eq:kappa2def}
\end{equation}
$\kappa^2 < 1$ [$0\leq \lambda < (1-r/R_0)$] indicating circulating particles, while trapped particles have $\kappa^2 >1$ [$(1-r/R_0) <  \lambda \leq (1+r/R_0)$]. Using the $(s,\alpha)$ model tokamak equilibrium~\cite{connor78} ($\alpha=-R_0q^2d\beta/dr$), the following expressions for the (transit, bounce) time of (circulating, trapped) particles are obtained:
\begin{equation}
\tau_b^{-1} =  \left( \frac{1}{ 4 {\rm I} \! {\rm K} (\kappa) } \, , \, \frac{\kappa}{4 {\rm I} \! {\rm K} (1/\kappa) } \right) \frac{\left( 2 {\cal E} \right)^{1/2}}{qR_0}
\left[ \frac{2(r/R_0)}{2(r/R_0) + (1-r/R_0) \kappa^2}\right]^{1/2} \;\; \label{eq:taub}
\end{equation}
Here, ${\rm I} \! {\rm K}$ stands for the complete elliptic integral of the first kind. In the same way, the bounce averaged precession frequency $\bar{\omega}_d$ can be computed as~\cite{rosenbluth71,connor83}:
\begin{eqnarray}
\hspace*{-2.4cm} \bar{\omega}_d&=&  \frac{{\cal E}}{\omega_{c} R_0}\frac{(q/r)(\kappa^2+4r/R_0)}{2(r/R_0)+(1-r/R_0)\kappa^2}
\left[  1  + \frac{2}{\kappa^2} \left( \frac{{\rm I} \! {\rm E} (\kappa)}{{\rm I} \! {\rm K} (\kappa)} -1 \right) 
- \frac{4\alpha}{3\kappa^2} \left( 2 (1-1/\kappa^2) 
 \right. \right. \nonumber \\
\hspace*{-2.4cm} & & + \left. (2/\kappa^2-1) \frac{{\rm I} \! {\rm E} (\kappa)}{{\rm I} \! {\rm K} (\kappa)} \right)  - \frac{\kappa^2}{\kappa^2+4r/R_0}\frac{\alpha}{2q^2} \left. + \frac{4}{\kappa^2}  s \left( \frac{{\rm I} \! {\rm E} (\kappa)}{{\rm I} \! {\rm K} (\kappa)} - \frac{\pi}{2 {\rm I} \! {\rm K} (\kappa)} \left( 1 - \kappa^2 \right)^{1/2} \right)\right] \label{eq:omegadcirc}
\end{eqnarray}
for circulating particles ($\kappa^2 < 1$), whereas, for magnetically trapped particles ($\kappa^2 > 1$)~\cite{graves00,rosenbluth71,connor83},
\begin{eqnarray}
\bar{\omega}_d&=& \frac{{\cal E}}{\omega_{c} R_0} \frac{q}{r}
\left[   \frac{2 {\rm I} \! {\rm E} (1/\kappa)}{{\rm I} \! {\rm K} (1/\kappa)} -1  + 4  s \left( \frac{{\rm I} \! {\rm E} (1/\kappa)}{{\rm I} \! {\rm K} (1/\kappa)} + \frac{1}{\kappa^2} - 1\right)
 \right. \nonumber \\
& &\left. - \frac{\alpha}{2q^2} - \frac{4\alpha}{3} \left(  1-1/\kappa^2 + (2/\kappa^2-1) \frac{{\rm I} \! {\rm E} (1/\kappa)}{{\rm I} \! {\rm K} (1/\kappa)} \right) \right] \;\; , \label{eq:omegadtrap}
\end{eqnarray}
where  ${\rm I} \! {\rm E}$ stands for the complete elliptic integral of the second kind. By direct inspection of Eqs.~(\ref{eq:taub}) to~(\ref{eq:omegadtrap}) and accounting for the fact that $\int d {\cal E} d \lambda = \int d {\cal E} d \kappa^2 (2r/R_0) \left[ 2(r/R_0)+(1-r/R_0)\kappa^2 \right]^{-2}$ by definition of $\kappa^2$, we see that only circulating particles with $(r/R_0)^{1/2} \laeq \kappa^2 <1$ contribute to $\delta \hat{W}_k$ on the same footing as trapped particles with $\kappa^2 >1$. Meanwhile, $\kappa^2$ is the strength of the poloidal modulation of the parallel velocity along the particle trajectory; thus, we denominate circulating particles with $(r/R_0)^{1/2} \laeq \kappa^2 <1$ as {\em barely circulating} to distinguish them from the {\em well circulating} particles with $\kappa^2<(r/R_0)^{1/2}$. The peculiar roles of trapped and barely circulating particles will be further discussed in Section~\ref{sec:inertia} in connection with the generalized inertia term, $\Lambda$, appearing in Eqs.~(\ref{eq:fish0}) and~(\ref{eq:fish1}).

Equations~(\ref{eq:fish0}) and~(\ref{eq:fish1})
generalize the electron fishbone dispersion relations, analyzed recently~\cite{zhou05,wang06,sun05}, to both trapped and barely circulating fast particles, including $(s,\alpha)$ model equilibrium effects on $\bar{\omega}_d$. A detailed discussion of the circulating electron effect on $\delta \hat{W}_k$ was recently given in \cite{wang06b}. A further extension of Eqs.~(\ref{eq:fish0}) and~(\ref{eq:fish1}) to a broader frequency range than that usually assumed near the ion diamagnetic gap~\cite{coppi86} in the low-frequency shear Alfv\'{e}n continuum is discussed in Section~\ref{sec:inertia}.

\subsection{Generalized inertia and high-frequency fishbones}
\label{sec:inertia}

For the present scope, we need an explicit expression of the generalized inertia term, $\Lambda$, appearing in Eqs.~(\ref{eq:fish0}) and~(\ref{eq:fish1}), for two limiting cases: (i) the banana regime, $|\omega|\ll \omega_{bi} \ll \omega_{ti}$, with $\omega_{bi}(\omega_{ti})$ the thermal ion bounce(transit) frequency, where~\cite{graves00,mikhailovskii83}
\begin{equation}
 \Lambda^2 = \left( \omega^2/\omega_A^2\right)\left(1-\omega_{*pi}/\omega \right) \left[1 + \left( 1.6 (R_0/r)^{1/2} + 0.5 \right) q^2 \right]\;\; ; \label{eq:lambdalow}
\end{equation}
(ii) the high frequency regime, $|\omega|\gg \omega_{ti}$, where~\cite{zonca96}
\begin{equation}
\hspace*{-1.2cm}\Lambda^2 = \frac{\omega^2}{\omega_A^2}- \frac{\omega_{BAE}^2}{\omega_A^2} \left[ 1 + \frac{\omega_{BAE}^2}{q^2 \omega^2} \frac{(46/49)+ (32/49) (T_e/T_i) + (8/49) (T_e/T_i)^2}{\left(1+(4/7)(T_e/T_i)\right)^2}\right]\; . \label{eq:lambdahigh}
\end{equation}
Here, $\omega_A=v_A/(qR_0)$, $v_A$ is the Alfv\'{e}n speed, $\omega_{*pi}={\bf k} \times {\bf B}/B \cdot {\bf \nabla} P_i/(n_i m_i \omega_{ci})$, ${\bf k}$ is the  wave-vector, $\omega_{BAE}=q\omega_{ti} (7/4 + T_e/T_i)^{1/2}$ and $\omega_{ti}=(2T_i/m_i)^{1/2}/(qR_0)$. 
The shear Alfv\'{e}n frequency gap is given by the condition ${\rm I} \! {\rm Re} \Lambda^2 < 0$~\cite{zonca06,chen06} ($\Lambda$ is generally complex), while the shear Alfv\'{e}n continuous spectrum is described by~\cite{zonca96} 
\begin{equation}
 \Lambda^2 = k_\|^2 q^2 R_0^2 \;\; . \label{eq:cont}
\end{equation}

The correct form of the enhancement factor $\propto q^2$ in Eq.~(\ref{eq:lambdalow}) was first pointed out in \cite{graves00}: the $1.6 (R_0/r)^{1/2} q^2$ factor comes from trapped $\kappa^2>1$, 
and barely circulating particles, $1> \kappa^2 \gaeq (r/R_0)^{1/2}$; 
the $0.5 q^2$ term, meanwhile, is due to well circulating particles, $\kappa^2\laeq (r/R_0)^{1/2}$ (see \ref{app:inertia} for a more detailed discussion). It differs from the well known $2q^2$ factor~\cite{glasser75} due to the intrinsic limitation of the ideal MHD model in assuming an isotropic pressure response: $2q^2$ would be the result for $\delta P = \delta P_\parallel$, while $\delta P_\perp \neq \delta P_\parallel$ for the geodesic curvature dynamics in toroidal systems. The problem of the kinetic {\em bulk ion inertia enhancement} for low frequency (banana-regime) MHD modes was analyzed in Refs.~\cite{mikhailovskii83,mikhailovskii79,belikov92}, where estimates were given for both inertia enhancement as well as ion Landau damping. A more systematic analytic approach was given in Refs.~\cite{graves00} and~\cite{bondeson96}.  More recently, it was pointed out that ion Landau damping due to the precession resonance with thermal (bulk) ions may be of crucial importance in determining the internal kink mode stability in ITER~\cite{hu06}. Here, it is worthwhile noticing that the inertia enhancement factor is identical to the zonal flow (ZF) polarizability induced by Ion Temperature Gradient (ITG) turbulence~\cite{rosenbluth98,hinton99}. This is not a coincidence and is due to the fact that, at long wavelengths, shear Alfv\'{e}n wave compressibility due to geodesic curvature coupling at $k_\|=0$ is identical to the corresponding dynamics of electrostatic waves with $k_\zeta=k_\theta=0$, provided that diamagnetic effects are neglected. For this reason, we must expect a correspondence between ZF polarizability and shear Alfv\'{e}n wave inertia enhancement in the banana regime, as in Eq.~(\ref{eq:lambdalow}); a similar correspondence is expected between GAM and Eq.~(\ref{eq:lambdahigh}), as pointed out in \cite{chen06,zonca06b} (see also the following discussion). 

Similar considerations apply for Eq.~(\ref{eq:lambdahigh}), for $\omega_{ti}\ll |\omega| \ll \omega_A$, where the $\propto 1/q^2$ term is different from $(2q^2)^{-1}$, predicted by ideal MHD~\cite{winsor68}. 
It was proposed in \cite{mikhailovskii73}, within the limits of a local approximation (i.e. without the proof of
the existence of unstable eigenmodes), that compressibility effects, associated with wave-particle
resonances due to the periodic toroidal transit motion of thermal ions, may be a
source of instability for short wavelength shear Alfv\'{e}n waves. Later, other authors~\cite{hastie81,cheng82} numerically demonstrated the
existence, well below the ideal stability threshold, of electromagnetic instabilities due to ion magnetic
drift resonances, assuming the very short wavelength limit $|\omega_{ti}|\ll |\omega| \approx |\omega_{di}|$. The effect of ion transit resonances was reconsidered in \cite{kotschenreuther86,zheng94}, where it was demonstrated numerically that the 
$\omega=\omega_{ti}$ resonance has analogous
effects to those of $\omega=\omega_{di}$, and in \cite{romanelli91}, where these effects on resistive interchange modes
were analyzed. All these analyses of short
wavelength drift-type modes are important for the present investigation since
the inertial (kinetic) layer physics is the same at high and low mode numbers~\cite{pegoraro86} and, therefore, they are relevant for the computation of the renormalized plasma inertia for low frequency MHD fluctuations.
For the same reason, these studies were readily
extended to the investigation of long-wavelength (low-mode-number) MHD modes~\cite{bondeson96,romanelli91b,zheng96}.
With the same expression of $\Lambda$, derived in Refs.~\cite{mikhailovskii73,kotschenreuther86}, Ref.~\cite{zonca96} demonstrated the existence conditions of fluctuations of the shear Alfv\'{e}n branch, excited by both energetic as well as thermal ions below the ideal MHD stability threshold, based on the general fishbone-like dispersion relation~\cite{zonca06,chen06} in the form of Eq.~(\ref{eq:fish0}). 
In the long wavelength limit, the expression of $\Lambda$ of Refs.~\cite{zonca96,mikhailovskii73,kotschenreuther86} accounts for the inertia enhancement as well as ion Landau damping for $\omega_{bi} \ll |\omega| \ll \omega_A$. For $\omega_{ti} \ll |\omega|$, it reduces to Eq.~(\ref{eq:lambdahigh}) (see also \cite{bondeson96}), with an exponentially small ion Landau damping, $\propto \exp ( -\omega^2/\omega_{ti}^2 )$. This favors the formation of fishbone gap modes near the BAE accumulation point for conditions with $\omega_{BAE} \gg \omega_{ti}$, i.e. $T_e/T_i \gg 1$ and/or $q\gg 1$~\cite{zonca_easter03}. Note that, due to the symmetry of the frequency gap described by Eq.~(\ref{eq:lambdahigh}), fishbone gap modes near the BAE accumulation point can be equally excited in both ion as well as electron diamagnetic directions. 
Meanwhile, the existence condition for the ``BAE"-fishbone gap mode is given by ${\rm I} \! {\rm Re} \left( \delta \hat{W}_f + \delta \hat{W}_k \right) < 0$~\cite{zonca06,chen06}.
That the shear Alfv\'{e}n continuum accumulation point ($\Lambda^2=0$) given by Eq.~(\ref{eq:lambdahigh}) is degenerate with the GAM frequency~\cite{winsor68}, as pointed out in \cite{chen06,zonca06b}, can be verified by direct comparison with the kinetic expression of the GAM frequency given by Ref.~\cite{lebedev96}. The degeneracy of BAE accumulation point and GAM frequency has been recently noted also in Ref.~\cite{garbet06}.

\section{Linear excitation of electron fishbones}
\label{sec:linear}

In this Section, we examine more closely the excitation of electron-fishbones on the basis of the mode dispersion relations, Eqs.~(\ref{eq:fish0}) and~(\ref{eq:fish1}), introduced and analyzed in Section~\ref{sec:disprel}. We also discuss some experimental evidence of both low- as well as high-frequency fishbones, for which the generalized inertia term is given by Eqs.(\ref{eq:lambdalow}) and~(\ref{eq:lambdahigh}), respectively. This frequency classification strictly applies to discrete gap modes, which tend to be excited nearby the shear Alfv\'{e}n continuum accumulation points. It can be extended to continuum resonant modes as well, when the mode drive is sufficiently weak that proximity to accumulation points matters for minimizing continuum damping. Generally, strongly driven continuum resonant modes can be excited regardless the shear Alfv\'{e}n continuum structure.

\subsection{Low-frequency fishbones}
\label{sec:low}

The crucial features of low-frequency electron fishbone excitations are dictated by the asymmetry of the shear Alfv\'{e}n continuum structure at low frequency~\cite{coppi86}, quantitatively expressed by Eq.~(\ref{eq:lambdalow}), which favors the excitation of modes propagating in the ion diamagnetic direction. Consistently with experimental observations~\cite{wong00,ding02,li02}, high field side ECRH fulfills this requirement and guarantees both drift-reversal of the barely trapped supra-thermal electrons as well as the inverted spatial gradient of the supra-thermal tail ($\omega_{*}/\omega > 0$) necessary for effective mode excitation. The case of mode excitation by LH only~\cite{smeulders02,romanelli02} follows the same physics with few additional twists. The fast electron population which effectively excite the mode are the trapped and barely circulating particles ($\kappa^2 \gaeq (r/R_0)^{1/2}$), because of Eqs.~(\ref{eq:deltawk}) and~(\ref{eq:omegadcirc}). Meanwhile, LH power forms a parallel as well as a perpendicular fast electron tail (via Coulomb collisions), which is moderately slanted toward the counter-current direction; i.e., despite that it guarantees the inverted spatial gradient of the supra-thermal tail ($\omega_{*}/\omega > 0$), it is less selective than high field side ECRH in producing particles with drift-reversal. In the case of mode excitation by LH only~\cite{smeulders02,romanelli02}, the presence of circulating supra-thermal particles is crucial for two reasons: (i) barely circulating particles ($\kappa^2 \gaeq (r/R_0)^{1/2}$) effectively contribute to the mode excitation as described by Eq.~(\ref{eq:deltawk}); (ii) well circulating particles ($\kappa^2 \laeq (r/R_0)^{1/2}$) modify the current profile, eventually reversing the magnetic shear and broadening the fraction of trapped particles characterized by drift reversal, as shown in Eq.~(\ref{eq:omegadtrap}).
Note that this effect modifies directly the kinetic contribution to the internal kink potential energy and is not associated with the MHD (fluid) potential energy change, controlled by LH power via current profile modification, as recently discussed for explaining HT-7 observations~\cite{sun05b}.
As in the case of  ion fishbones, two branches of the electron fishbone exist: a discrete gap mode~\cite{coppi86} and a continuum resonant mode~\cite{chen84}. The latter does not generally require neither drift-reversal nor inverted spatial gradient of the supra-thermal tail; however, it has a higher excitation threshold and, thus, it is unfavored, particularly for the branch propagating in the $\omega_{*e}$ direction.

Applying Eq.~(\ref{eq:fish1}) to FTU shot \# 20865 (see Figure~1), %Figure~\ref{fig:figure1}), 
the almost steady oscillation of the mode in the low LH power phase and the absence of sawtooth oscillations suggest that $1\gg \Delta q_s >0$. This is consistent with the $q$-profile reconstruction by transport simulations, reported in Figure~2 %Figure~\ref{fig:figure2} 
(FTU has no $q$ profile measurements near the magnetic axis). Even in the high LH power phase (Figure~3), %(Figure~\ref{fig:figure3}), 
the minimum-$q$ value remains extremely near unity. From experimental observations, $\omega\simeq 60$ krad/s, $\omega_{*pi}\simeq 23$ krad/s, $\omega_{bi}\simeq 70$ krad/s, $\omega_{ti}\simeq 400$ krad/s, $\omega_{BAE}\simeq 900$ krad/s and $\omega_A \simeq 9.5$ Mrad/s. Thus $\omega_{*pi}<\omega\laeq\omega_{bi}\ll\omega_{ti}$ and we can apply Eq.~(\ref{eq:lambdalow}), showing $\Lambda^2>0$. Given the $\omega\laeq\omega_{bi}$ condition, a further generalization of Eq.~(\ref{eq:lambdalow}) would be necessary for a rigorous analysis including mode damping by precession~\cite{hu06} and precession-bounce resonances with thermal ions. These results, however, would simply lead to a redefinition of the mode excitation threshold (see Eq.~(\ref{eq:betahc}) below) at the expense of technical complications; thus, they will be reported elsewhere. 
\begin{figure}
\begin{center}
\noindent\epsfxsize=0.75\linewidth\epsfbox{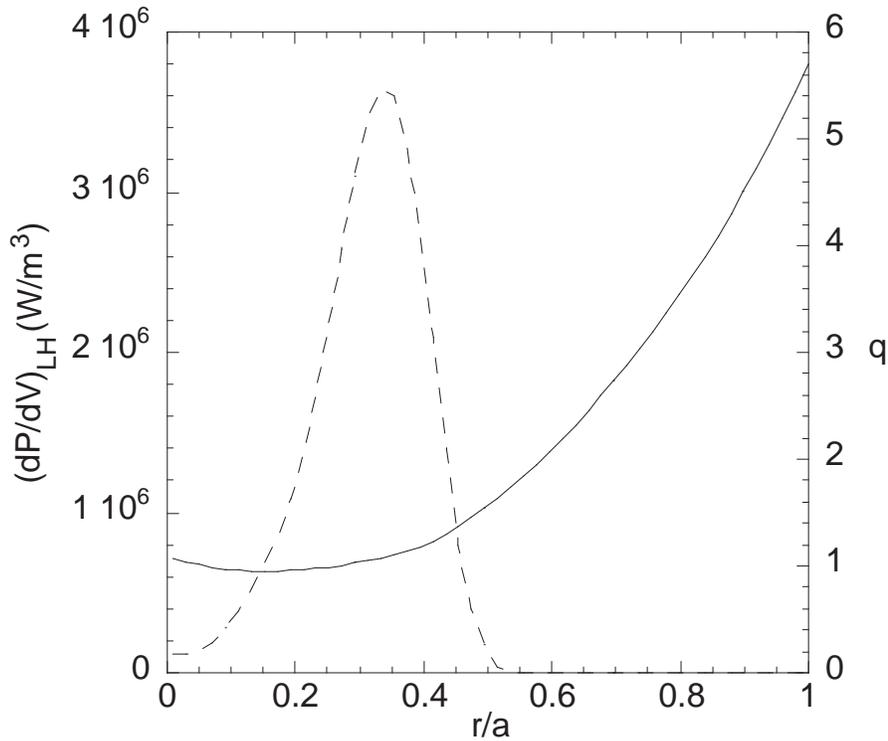}
\end{center}
\label{fig:figure2}
\caption{Absorbed LH power density (broken line) and $q$ profile (solid line) at $t=220$ ms, as predicted from transport simulations of FTU shot \# 20865. The total absorbed LH power is $P_{LH}=0.76$ MW.}  
\end{figure}
\begin{figure}
\begin{center}
\noindent\epsfxsize=0.75\linewidth\epsfbox{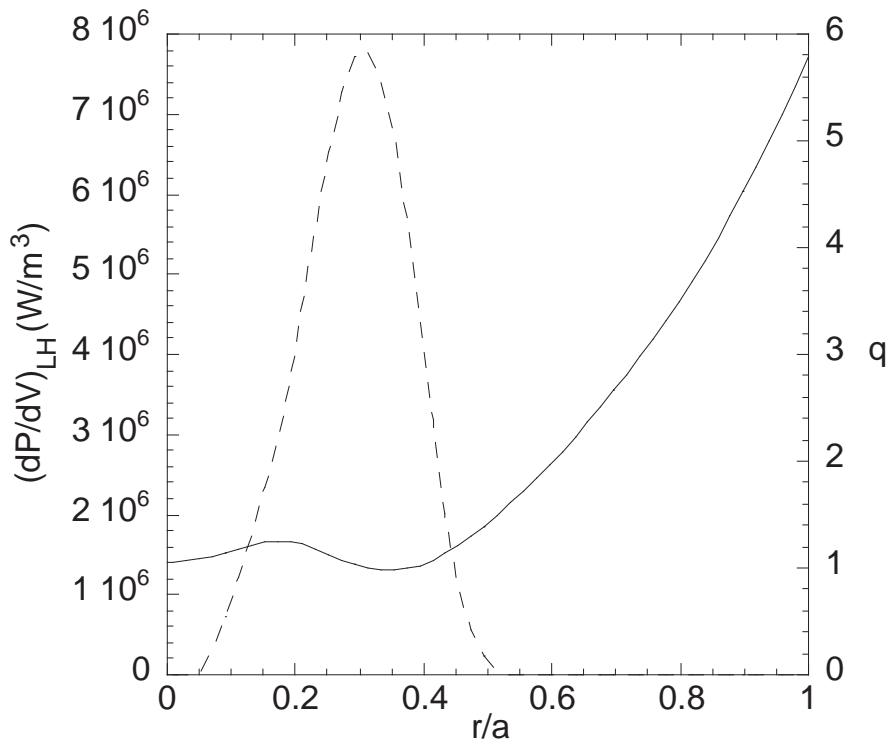}
\end{center}
\label{fig:figure3}
\caption{Absorbed LH power density (broken line) and $q$ profile (solid line) at $t=280$ ms, as predicted from transport simulations of FTU shot \# 20865. The total absorbed LH power is $P_{LH}=1.69$ MW.}  
\end{figure}
Given Eq.~(\ref{eq:fish1}), for $\Lambda^2> \Delta q_s^2$ the mode can be considered as continuum resonant mode~\cite{chen84}, following the standard classification~\cite{chen84}. Meanwhile,  $\Lambda^2< \Delta q_s^2$ would correspond to a gap mode~\cite{coppi86}. In either case we can assume $\Lambda^2 \sim \Delta q_s^2 = {\rm O}(10^{-4})$, consistent with $q$-profile reconstruction by transport simulations and with experimental observations. At larger values of $\Delta q_s$, the mode frequency near the accumulation point would rapidly increase up to the BAE frequency and Eq.~(\ref{eq:lambdahigh}) would apply, rather than Eq.~(\ref{eq:lambdalow}). Besides the obvious consequence of increasing the MHD stability of the system, i.e. $\delta \hat{W}_f$, this fact would imply that higher effective supra-thermal electron temperature are needed for both balancing $\delta\hat{W}_f$ by $\delta{\rm I}\! {\rm Re}\hat{W}_k$ and for efficiently driving the mode via wave particle resonances (see Section~\ref{sec:high}). Altogether, we expect that increasing $\Delta q_s$ increases the stability of the system, as verified experimentally on FTU. 

In the case of the gap mode near the accumulation point~\cite{coppi86}, the existence condition is $\delta \hat{W}_f + \delta {\rm I}\! {\rm Re}\hat{W}_k < 0$ and real mode frequency is given by
\begin{equation}
 \left({\rm I}\! {\rm Re}\Lambda\right)^2 = \Delta q_s^2 - \frac{\left( \delta \hat{W}_f + {\rm I}\! {\rm Re} \delta \hat{W}_k\right)^2}{S^2 \Delta q_s} \;\; , \label{eq:omegagap}
\end{equation}
while the growth rate is obtained from
\begin{equation}
 \gamma = \Gamma \left( g {\rm I}\! {\rm Im} \delta \hat{W}_k - {\rm I}\! {\rm Im} \Lambda \right) \;\; , \label{eq:gammagap}
\end{equation}
where $g\equiv - \left( \delta \hat{W}_f + {\rm I}\! {\rm Re} \delta \hat{W}_k\right)/\left( S^2 \Delta q_s {\rm I}\! {\rm Re}\Lambda \right)$ and $\Gamma^{-1}=\partial {\rm I}\! {\rm Re}\Lambda/ \partial \omega - g \partial {\rm I}\! {\rm Re}\delta \hat{W}_k/ \partial \omega$.

For the continuum resonant mode~\cite{chen84}, Eq.~(\ref{eq:fish1}) can be written as 
\begin{equation}
 {\rm i} S \left( \Lambda^2 - \Delta q_s^2 \right)^{1/2} \left[ \Delta q_s - {\rm i} \left( \Lambda^2 - \Delta q_s^2 \right)^{1/2} \right]^{1/2} = \delta \hat {W}_f + \delta \hat {W}_k \;\;  .  \label{eq:fish2}
\end{equation}
Assuming $\Delta q \rightarrow 0$, for simplicity, 
the mode dispersion relation becomes
\begin{equation}
 \delta \hat {W}_f + {\rm I} \! {\rm Re} \delta \hat {W}_k = (S/\sqrt{2}) \Lambda^{3/2} \simeq 0 \;\; , \label{eq:fishre}
\end{equation}
which determines the mode frequency~\cite{chen84}; meanwhile, the mode growth rate is defined by~\cite{white90}
\begin{equation}
 \gamma = \Gamma \left[ \int_0^{r_s} \left(r/r_s\right) \left( \partial \beta_{h,res}/\partial r \right) dr - \beta_{h,c} \right] \;\; , \label{eq:fishim}
\end{equation}
where $\Gamma= - (R_0/r_s) (\partial {\rm I} \! {\rm Re} \delta \hat {W}_k / \partial \omega - 3S/(2\sqrt{2}) \Lambda^{1/2}\partial \Lambda /\partial \omega)^{-1}$, the effective resonant fast electron normalized pressure, $\beta_{h,res}$, is defined such that ${\rm I} \! {\rm Im} \delta \hat{W}_k \equiv (R_0/r_s^2) \int_0^{r_s} r dr \partial_r  \beta_{h,res}$ and the critical excitation threshold $\beta_{h,c}$ is given by
\begin{equation}
 \beta_{h,c}= (r_s/R_0) (S/\sqrt{2}) \Lambda^{3/2} \;\; . \label{eq:betahc}
\end{equation}
Note that the $\propto \beta_{h,res}$ term in Eq.~(\ref{eq:fishim}) would change sign for the case of mode excitations by fast ions.

Despite the different structures of Eqs.~(\ref{eq:omegagap}) and~(\ref{eq:gammagap}) with respect to Eqs.~(\ref{eq:fishre}) and~(\ref{eq:fishim}), their extension to the nonlinear regime follows the same derivation. For this reason,   we derive the nonlinear amplitude equations describing the fishbone cycle, in Section~\ref{sec:NL}, limiting specific applications to the simple case of Eqs.~(\ref{eq:fishre}) and~(\ref{eq:fishim}). Analogous derivations in other more general cases, included in Eqs.~(\ref{eq:fish0}) and~(\ref{eq:fish1}), follow consequently. 

With FTU shot \# 20865 data, $S=0.52$ and $\beta_{h,c} \simeq 0.34 \times 10^{-4}$ at $t=220$ ms, while $S=4.4$ and $\beta_{h,c} \simeq 4.9 \times 10^{-4}$ at $t=280$ ms. Lower Hybrid power deposition computations provide the supra-thermal electron tail distribution function~\cite{cardinali_easter03}, which give $\int_0^{r_s} (r/r_s) dr \partial_r  \beta_{h,res} \simeq 0.85 \times 10^{-4}$ at $t=220$ ms (see Figure~4) %(see Figure~\ref{fig:figure4}) 
and $\int_0^{r_s} (r/r_s) dr \partial_r  \beta_{h,res} \simeq 6.0 \times 10^{-4}$ at $t=280$ ms (see Figure~5). %(see Figure~\ref{fig:figure5}). 
This is consistent with mode excitations and the transition observed in Figure~1, %Figure~\ref{fig:figure1}, 
when the power level is stepped from given $P_{LH}=0.76$ MW up to $P_{LH}=1.69$ MW. The bursting fishbone activity is further discussed in Section~\ref{sec:NL}.
\begin{figure}
\begin{center}
\noindent\epsfxsize=0.75\linewidth\epsfbox{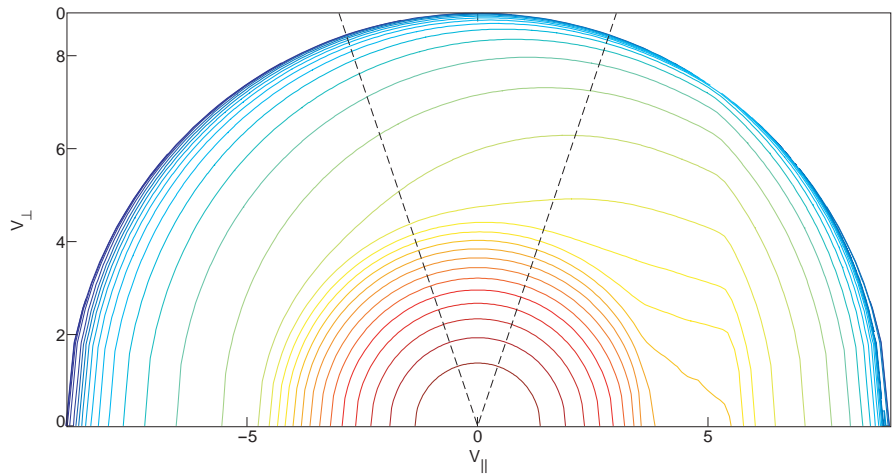}
\end{center}
\label{fig:figure4}
\caption{Contour plot of the supra-thermal electron tail at $t=220$ ms, as predicted from Fokker-Planck computations of FTU shot \# 20865. The total absorbed LH power is $P_{LH}=0.76$ MW. Velocities are normalized to the core electron thermal speed. The radial position is $r/a=0.17$ and dashed lines indicate the trapped particle region.}  
\end{figure}
\begin{figure}
\begin{center}
\noindent\epsfxsize=0.75\linewidth\epsfbox{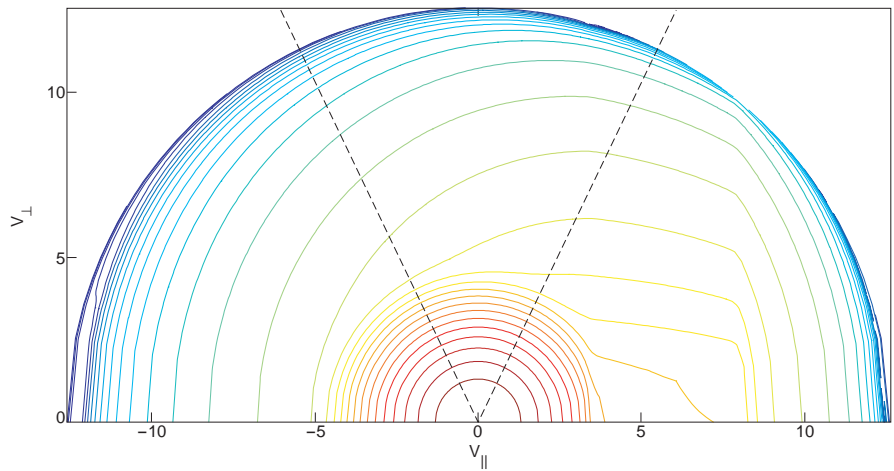}
\end{center}
\label{fig:figure5}
\caption{Contour plot of the supra-thermal electron tail at $t=280$ ms, as predicted from Fokker-Planck computations of FTU shot \# 20865. The total absorbed LH power is $P_{LH}=1.69$ MW. Velocities are normalized to the core electron thermal speed. The radial position is $r/a=0.35$ and dashed lines indicate the trapped particle region.}  
\end{figure}

\subsection{High-frequency fishbones}
\label{sec:high}

At higher frequencies, $\omega\gg \omega_{ti}$, Eq.~(\ref{eq:lambdahigh}) applies instead of Eq.~(\ref{eq:lambdalow}); thus, the asymmetry of the shear Alfv\'{e}n continuous spectrum is lost and modes can equally propagate in both ion and electron diamagnetic directions. Equation~(\ref{eq:lambdahigh}) describes the formation of the Beta induced Alfv\'{e}n Eigenmodes (BAE)~\cite{heidbrink93} spectral gap: so, electron fishbones propagating in the electron diamagnetic direction and normal pressure profiles could be excited.
More precisely, high power ECRH experiments with on axis resonance would be needed, producing sufficiently high effective supra-thermal electron tail temperatures, $T_h$, for the fast particle precession frequency to be of the order of the thermal ion transit frequency.
For the above FTU parameters, this would require $T_h\gaeq 100$ keV, to be compared with the usual values $T_h\simeq 30$ keV, as well as $T_e\gg T_i$ for consistency (see Section~\ref{sec:inertia}). Obviously, at such high energies of the supra-thermal electron tail, relativistic effects can be important and should be included in the expression of $\delta \hat{W}_k$~\cite{wang06}. 

The existence condition of gap modes in the BAE frequency gap just below the continuum accumulation point is given by ${\rm I} \! {\rm Re} \left( \delta \hat{W}_f + \delta \hat{W}_k \right)<0$~\cite{zonca06,chen06}, as discussed in Section~\ref{sec:inertia}.
Note that these fishbones, possibly excited below the BAE frequency, could be equally excited by ICRH induced fast ions but, in that case, they would propagate in the ion diamagnetic direction. The observation of high frequency precessional fishbones with ICRH in JET~\cite{nabais04,nabais05} can be possibly interpreted as evidence of fishbone excitation below the BAE frequency, as predicted by theory. 
One striking evidence that Eqs.~(\ref{eq:fish0}) and~(\ref{eq:lambdahigh}) describe these physics is Fig.~11 of \cite{nabais05}. In fact, as the diamagnetic fishbone get excited and less free energy is available for the excitation of the precessional fishbone (modes are less strongly driven), theory predicts that frequency chirping should decrease and the mode frequency should get closer to the accumulation point. However, this is evidently not the usual accumulation point at $\omega_{* pi}$, but rather the accumulation point described by Eq.~(\ref{eq:lambdahigh}). In fact, Fig.~11 of \cite{nabais05} shows the frequency accumulation at about 70 kHz. To test this conjecture, we have computed the BAE accumulation point in two ways: (a) via the simplified expression $\omega_{BAE}=q\omega_{ti} (7/4+T_e/T_i)^{1/2}$; and (b) via numerical solution of $\Lambda=0$, with $\Lambda$ given by Ref.~\cite{zonca96}, i.e. including both thermal ion transit resonances (for the ion Landau damping evaluation) as well as diamagnetic effects (finite $\omega_{* pi}$).
For the JET discharge \# 54300 ($D$ plasma with ICRH $H$-minority heating), we have taken $T_e=6$ keV, $R_0=3$ m and $\eta_i=\partial \ln T_i/\partial \ln n=2$, obtaining the results reported in Table~\ref{tab:th}.
\begin{table}
\begin{center}
\begin{tabular}{|c|c|c|c|c|c|c|} \hline \hline 
$\omega_{\Lambda=0}/\omega_{ti}$   &  $\gamma/\omega_{ti}$ &  $\omega_{*ni}/\omega_{ti}$ & $T_i$ & $\omega_{\Lambda=0}/2\pi $  & $\omega_{*pi}/2\pi$ & $\omega_{BAE}/2\pi $ \\ \hline \hline
.24100E+01   &   -.07500E-01   &     .10000E+00    &    3 keV  &  68 kHz &  8.5 kHz  & 55  kHz  \\ \hline
.22900E+01   &   -.06700E-01   &     .15000E+00    &    3 keV  &  65 kHz  &  13 kHz  & 55  kHz  \\ \hline
.21700E+01   &   -.03200E-01   &     .20000E+00    &    3 keV  &  62 kHz &  17 kHz &  55  kHz \\ \hline 
.22900E+01   &   -.10700E-01   &     .10000E+00    &    4 keV  &  75 kHz  & 9.9 kHz   & 59 kHz  \\ \hline
.21900E+01   &   -.09700E-01   &     .15000E+00    &    4 keV  & 72 kHz   &  15 kHz  & 59 kHz  \\ \hline
.20700E+01   &   -.06300E-01   &     .20000E+00    &    4 keV  &   68 kHz &  20 kHz &  59 kHz \\ \hline
\end{tabular}
\end{center}
\caption{\label{tab:th} Theoretical values of the BAE accumulation point, $\omega_{\Lambda=0}$ from $\Lambda=0$~\cite{zonca96}, as a function of $T_i$ and $\omega_{*ni}=\omega_{*pi}/(1+\eta_i)$. Fixed parameters are $T_e=6$ keV, $R_0=3$ m and $\eta_i=2$. Values of ion Landau damping, $\gamma$, are also reported.}
\end{table}
Values of Landau damping are typically small. Meanwhile, comparisons of theoretical frequencies with the experimental value of $\simeq 70$ kHz suggest that a realistic estimate for $T_i$ at the $q=1$ surface is $T_i\simeq 4$ keV with $15$ kHz $\laeq \omega_{*pi}\laeq 20$ kHz, in agreement with experimental observations~\cite{nabais05}.
The crystal spectrometer for this case gives $T_i=2.2$ keV, which is a lower bound of the ion temperature at the $q=1$ surface and approximately $60\%$ of its value, as suggested by normal experience. The good agreement of theoretical predictions with experimental observations confirms the sound basis of the proposed interpretation of high frequency precessional fishbones observed in JET~\cite{nabais04,nabais05} with ICRH as evidence of fishbone excitation below the BAE frequency~\cite{zonca_easter03}.
The scaling of the BAE accumulation point frequency with $T_e/T_i$ can be used for diagnostics purposes, similar to the approach proposed in Ref.~\cite{breizman05} for Alfv\'{e}n Cascades. Actually, the results presented here (see Table~\ref{tab:th}) and their dependence on diamagnetic effects show that a better evaluation of the accumulation point frequency can be obtained by solving $\Lambda=0$~\cite{zonca96} rather than using $\omega=\omega_{BAE}$~\cite{breizman05}, with the additional advantage of computing ion Landau damping. In the case of Alv\'{e}n Cascades, of course, the accumulation point at $s=0$ should be evaluated using $\Lambda^2=k_{\| s}^2 q_s^2 R_0^2$~\cite{zonca96}, as predicted by Eq.~(\ref{eq:cont}). Note that magnetic shear never enters in the accumulation point expression, as expected for local oscillations of the shear Alfv\'{e}n continuum and explicitly shown by Eqs.~(\ref{eq:alpha12}) and~(\ref{eq:fish1}).

\section{Nonlinear amplitude equation}
\label{sec:NL}

FTU experimental results (see Figure~\ref{fig:figure1}) suggest that the level of LH power input controls the transition from nearly steady state to bursting electron fishbone oscillations. Here, we want to focus on the bursting regime, where we conjecture that the fishbone is a continuum resonant mode~\cite{chen84}, described by Eqs.~(\ref{eq:fishre}) and~(\ref{eq:fishim}), on the basis of preliminary high time resolution Electron Cyclotron Emission (ECE) measurements, which indicate that the frequency of temperature oscillations in the bursting phase drops by $40\div50 \%$ in $2\div3$ ms. Better resolved data on the mode frequency chirping are needed for more accurate comparisons of theory (this Section) with experiments. However, preliminary analyses support our conjecture that the bursting mode phase is associated with the excitation of a continuum resonant mode well above marginal stability, i.e. $|\beta_{h,res \, {\rm max}}-\beta_{h,res \, {\rm min}}| \sim \beta_{h,c}$ in Eq.~(\ref{eq:fishim}). For this reason, we expect that particle nonlinearities are dominant in dictating the time evolution of the fishbone cycle, as recently shown in Ref.~\cite{fu06}. The role of mode-mode couplings on the fishbone dynamics was specifically discussed in Ref.~\cite{odblom02}.

For strongly driven fishbone modes, fast particle dynamics is secular in the radial direction due to the
{\em mode-particle pumping} process, originally proposed in Ref.~\cite{white83}. Since, in this case, there is no time for the particles to experience trapping in the potential well of the wave, we use a different approach with respect to that of Ref.~\cite{berk97}, which postulates proximity to marginal stability and describes the nonlinear evolutions of modes with slowly varying frequencies due to structures in phase space near particle resonances. Here, we adopt the  4-wave modulation interaction model, introduced by Chen \etal~\cite{chen00} for analyzing modulational instabilities of the radial envelope of Ion Temperature Gradient driven modes in toroidal geometry, extending it to the modulations on the fast particle distribution function due to nonlinear mode dynamics, as proposed in Ref.~\cite{zonca00}. In the following, we show that the resonant particle motion is secular with a time-scale inversely proportional to the mode amplitude. In order to qualitatively compare the model predictions with FTU experimental results on the fishbone repetition rate, 
we show that our nonlinear model equations are expressible in terms of a predator-prey like model with a limit cycle. This model differs from the existing qualitative models (Refs.~\cite{chen84} and~\cite{coppi86}) in that it is structurally stable, i.e., the periodic dynamics not destroyed in the presence of higher-order perturbation terms (see \ref{app:predator}). The transition to the stable limit cycle behavior occurs via a marginal oscillatory regime (i.e., the {\em center} if to use the proper terminology), which is structurally unstable and is also revealed in those models discussed in Refs.~\cite{chen84} and~\cite{coppi86}.
%we show that our nonlinear model equations can be reduced to a predator-prey like system, which is derived in a structurally stable form (see \ref{app:predator}) that, for sufficient small oscillations about the fixed point, reduces to that of previously proposed phenomenological predator-prey models~\cite{chen84,coppi86}, which are structurally unstable.

We can generalize Eqs.~(\ref{eq:fish0}) and~(\ref{eq:fish1}) to include supra-thermal electron tail nonlinear dynamics by closely following the procedure of Ref. \cite{zonca05}. In the present treatment, as discussed above, we choose to neglect fishbone nonlinear dynamics associated with mode-mode couplings. For the case of continuum resonant fishbones~\cite{chen84}, this approximation allows us to retain the fundamental dynamics~\cite{chen06,fu06} and to make significant analytic progress, as shown below.

Under the action of the fishbone mode, the toroidally and poloidally symmetric (zonal~\cite{chen06,zonca06b}) nonlinear modification of the fast electron distribution function, Eq.~(\ref{eq:deltaf}), can be obtained  from the nonlinear gyrokinetic equation~\cite{frieman82} and is given by~\cite{zonca05}
\begin{equation}
\frac{\partial}{\partial t} H_{NL,z} = \sum_{{\bf k}_z= {\bf k}'+{\bf k}''} i \frac{c}{B_0} k_{\theta}' \frac{\partial}{\partial r} 
\overline{\left[ \left( 1 -\frac{k_{\parallel}'v_\parallel}{\omega_{k'}}  \right) \delta \phi_{k'} \delta H_{k''}\right]} \;\; ,  \label{eq:dhz0}
\end{equation}		
where $k_{\phi}'=-k_{\phi}''$, $k_{\theta}'=-k_{\theta}''$, we have neglected finite electron orbit widths and assumed $\delta E_{\parallel k}= 0$. By direct substitution and using Eqs.~(\ref{eq:deltaf}) and~(\ref{eq:deltak}), Eq.~(\ref{eq:dhz0}) is readily reduced to
\begin{equation}
\hspace*{-2.4cm}\frac{\partial}{\partial t} \delta H_{NL,z} = - \frac{2}{r} \omega_c \omega^2 \frac{\partial}{\partial r} \left[ \overline{e^{{\rm i}(1-q)\theta}\left( 1 - \frac{k_\| v_\parallel}{\omega} \right)} {\rm I} \! {\rm Im} \left( \frac{\overline{e^{{\rm i} q \theta} \omega_d e^{- {\rm i} \theta}}}{\bar{\omega}_d-\omega}\right) \left( \frac{QF_0}{\omega}\right) r^2 r_s^2 \left| \delta \xi_0\right|^2 \right]  . \label{eq:deltahnl}
\end{equation}
Here, $\overline{v_\parallel \exp {\rm i} (1-q) \theta}=0$ for trapped particles and $\delta \xi_0 = \delta \xi_{r0}/r_s$ is the normalized radial displacement of the mode, which is assumed to be the usual step function. The presence of the imaginary part of the particle response on the RHS of Eq.~(\ref{eq:deltahnl}) indicates the crucial roles played by resonant particles~\cite{chen99}. Meanwhile, by definition of the $Q F_0$ operator ($Q F_0/\omega \simeq \partial_{\cal E} F_0 + k_\theta/(\omega\omega_c) \partial_r F_0$), the RHS contains both $\propto \partial_r F_0$ and $\partial_r^2 F_0$ terms~\cite{chen99}. Thus, integrating both sides in velocity space, Eq.~(\ref{eq:deltahnl}) can be easily put in the form of a diffusion equation describing the relaxation of the fast particle profile within the $q=1$ surface:
\begin{equation}
 \frac{\partial}{\partial t} n_{h} = \dot{N}_{h} - \frac{2}{r} \omega_c \omega^2 \frac{\partial}{\partial r} \left[ r^2 r_s^2 \left| \delta \xi_0\right|^2 f_{eff,h} \left( \frac{Q_{res} n_{h}}{\omega}\right)\right] \;\; . \label{eq:ndiff}
\end{equation}
Here, $\dot{N}_{h}$ indicates the fast electron source term due to additional power input, we have defined the effective fraction of fast electrons $f_{eff,h}$ and 
\begin{equation}
\hspace*{-1.6cm} f_{eff,h} \left( \frac{Q_{res} n_{h}}{\omega}\right) = \left \langle F_0 \right \rangle^{-1} \left\langle \overline{e^{{\rm i}(1-q)\theta}\left( 1 - \frac{k_\| v_\parallel}{\omega} \right)} {\rm I} \! {\rm Im} \left( \frac{\overline{e^{{\rm i} q \theta} \omega_d e^{- {\rm i} \theta}}}{\bar{\omega}_d-\omega}\right) \left( \frac{QF_0}{\omega}\right) \right \rangle \; , \label{eq:feff}
\end{equation}
having indicated velocity space integration by angular brackets. From Eqs.~(\ref{eq:ndiff}) and~(\ref{eq:feff}) we recognize that the nonlinear diffusion coefficient due to the fishbone within the $q=1$ surface is given by $D_{NL}\simeq 2 \omega r_s^2 f_{eff,h} \left| \delta \xi_0\right|^2$. 

One obvious consequence of Eq.~(\ref{eq:deltahnl}) is the time evolution of the supra-thermal electron tail contribution to $\delta \hat{W}_f$ via their convective responses. In fact, nonlinearly~\cite{zonca05}
\begin{equation}
Q F_0 \rightarrow Q F_0 + \frac{k_\theta}{\omega_c} \frac{\partial}{\partial r}  \delta H_{NL,z} \;\; . \label{eq:nleq}
\end{equation}
Thus, the expression of $\partial_t \delta \hat{W}_{f,NL}$ is readily obtained from that of $\delta \hat{W}_f$ by direct substitution of the supra-thermal electron tail distribution, $\partial_r F_0$, with the expression of $\partial_t \partial_r \delta H_{NL,z}$ from Eq.~(\ref{eq:deltahnl}). For this reason,  in the present work we will simply assume it as given, without providing further detailed discussions.
The other effect of Eq.~(\ref{eq:deltahnl}) is to introduce a nonlinear modification to Eq.~(\ref{eq:deltak}) in the form
\begin{equation}
\hspace*{-1.2cm}\delta K_{NL} = \overline{\delta K}_{NL} e^{{\rm i} (\zeta - q(r)\theta )} = \frac{c}{B_0} \frac{\overline{e^{{\rm i} q(r)\theta}\omega_d e^{-{\rm i} \theta}}}{\bar{\omega}_d - \omega} \frac{k_\theta}{\omega} \frac{\partial \delta H_{NL,z} }{\partial r}  \delta \phi_0 (r) e^{{\rm i} (\zeta - q(r)\theta )} \;\; . \label{eq:deltaknl}
\end{equation}
Using Eq.~(\ref{eq:deltahnl}), meanwhile, the nonlinear modification for the resonant contribution (imaginary part) of $\delta\hat{W}_k$ is obtained in the form:
\begin{eqnarray}
& &\hspace*{-1.2cm}\left| \delta \xi_0\right|^{-2} \frac{\partial}{\partial t} \left[ \left(\frac{\partial}{\partial t} \delta \hat{W}_{k,NL}\right) \left| \delta \xi_0\right|^2 \right] \simeq - 8 i \frac{\pi^2}{B_0^2} m\omega_c^2 \omega \frac{R_0}{r_s^2} \int_0^{r_s} \frac{r^2}{q} dr \int {\cal E} d {\cal E} d\lambda  \nonumber \\
& & \hspace*{-0.6cm} \times \sum_{v_\parallel/|v_\parallel| = \pm 1} \tau_b \overline{e^{{\rm i} q(r)\theta}\omega_d e^{-{\rm i} \theta}}\, \overline{e^{{\rm i} \theta}\omega_d e^{-{\rm i} q(r) \theta}}  \frac{\partial}{\partial r} \left\{ k_\theta \frac{\partial}{\partial r} \left[ \overline{e^{{\rm i}(1-q)\theta}\left( 1 - \frac{k_\| v_\parallel}{\omega} \right)} \right. \right. \nonumber \\
& & \hspace*{-0.6cm} \times \left. \left. \overline{e^{{\rm i} q(r)\theta}\omega_d e^{-{\rm i} \theta}} {\rm I} \! {\rm Im} \left( \frac{QF_0}{\bar{\omega}_d - \omega} \right)  r^2 r_s^2 \left| \delta \xi_0\right|^2 \right] \right\} \;\; . \label{eq:deltawknl}
\end{eqnarray}
Note that here, as in the following Eqs.~(\ref{eq:deltawknlsimp}), (\ref{eq:dbetares}) and~(\ref{eq:redeltawknl}), partial time derivation is intended at constant frequency, which, in the present problem, can be itself a function of time and vary on the nonlinear time scale.
An intuitive derivation of Eq.~(\ref{eq:deltawknl}) can be obtained from Eqs.~(\ref{eq:deltawk}), (\ref{eq:deltak}) and~(\ref{eq:deltaknl}), noting that, for resonant particles involved in the $\delta H_{NL,z}$ dynamics
\begin{equation}
  \left( \bar{\omega}_d - \omega \right) \overline{\delta K}_{NL}  \simeq - \frac{\rm i} {\left| \delta \xi_0\right|} \frac{\partial}{\partial t} \left( \overline{\delta K}_{NL} \left| \delta \xi_0\right| \right) \;\; . \label{eq:intuitive}
\end{equation}
Recalling the definition of $\partial_r\beta_{h,res}$, given below Eq.~(\ref{eq:fishim}), i.e.
\begin{equation}
\hspace*{-2.4cm}\frac{\partial \beta_{h,res}}{\partial r}= 4 \frac{\pi^2}{B_0^2} m\omega_c^2 \frac{r^2}{q} \int {\cal E} d {\cal E} d\lambda \sum_{v_\parallel/|v_\parallel| = \pm 1}\overline{e^{{\rm i} q(r)\theta}\omega_d e^{-{\rm i} \theta}} \; \overline{e^{{\rm i} \theta }\omega_d e^{-{\rm i} q(r)\theta }} {\rm I} \! {\rm Im} \left(\frac{ \tau_b \, Q F_0}{\bar{\omega}_d - \omega} \right) \;\; , \label{eq:betahres}
\end{equation}
Eq.~(\ref{eq:deltawknl}) can be approximated as
\begin{equation}
\hspace*{-1.8cm}\frac{\partial}{\partial t}  \left[ \left(\frac{\partial}{\partial t} {\rm I} \! {\rm Im} \delta \hat{W}_{k,NL}\right) \left| \delta \xi_0\right|^2 \right] \simeq 2C \omega^2 R_0  \left| \delta \xi_0\right|^4 \int_0^{r_s} dr \frac{\partial}{\partial r} \left[ \frac{1}{r} \frac{\partial}{\partial r} \left( r^2 \frac{\partial}{\partial r} \beta_{h,res} \right) \right] \;\; . \label{eq:deltawknlsimp}
\end{equation}
Here, for simplicity, we have assumed that the radial variation of $\omega_d$ is essentially $\propto (1/r)$ within the $q=1$ (or minimum-$q$) surface; meanwhile, $C$ is a constant which may be computed exactly, given Eq.~(\ref{eq:deltawknl}) and the definition of $\beta_{h,res}$, Eq.~(\ref{eq:betahres}).
Using these results, the amplitude evolution equation can be formally written as Eq.~(\ref{eq:fishim}), 
\begin{equation}
(d/dt) \left| \delta \xi_0\right|^2 = 2 \Gamma \left[ \int_0^{r_s} \left(r/r_s\right) \left( \partial \beta_{h,res}/\partial r \right) dr - \beta_{h,c} \right] \left| \delta \xi_0\right|^2\;\; , \label{eq:fishimnl}
\end{equation}
where the nonlinear evolution equation for the resonant fast particle pressure gradient becomes
\begin{equation}
\hspace*{-1.8cm}\frac{\partial}{\partial t} \left[ \left| \delta \xi_0\right|^2 \left( \frac{\partial}{\partial t} - \nu_{ext} \right) \frac{\partial}{\partial r} \beta_{h,res} \right] = 2C \omega^2 \frac{r_s}{r}  \left| \delta \xi_0\right|^4 \frac{\partial}{\partial r} \left[ \frac{r_s}{r} \frac{\partial}{\partial r} \left( r^2 \frac{\partial}{\partial r} \beta_{h,res} \right) \right]\;\; . \label{eq:dbetares}
\end{equation}
Here, $\nu_{ext}$ is the reconstruction rate of $\beta_{h,res}$, i.e.,
\begin{eqnarray}
& &\nu_{ext}= 4 \frac{\pi^2}{B_0^2} \frac{m\omega_c}{\partial_r \beta_{h,res}} \frac{r^2 k_\theta }{q}  \int {\cal E} d {\cal E} d\lambda \sum_{v_\parallel/|v_\parallel| = \pm 1} \tau_b \overline{e^{{\rm i} q(r)\theta}\omega_d e^{-{\rm i} \theta}} \nonumber \\ & &\hspace*{4em}  \times {\rm I} \! {\rm Im} \left(  \frac{\overline{e^{{\rm i} \theta}\omega_d e^{-{\rm i} q(r) \theta}}}{\bar{\omega}_d - \omega} \right) \left( \frac{\partial}{\partial r} + \frac{\omega \omega_c}{k_\theta} \frac{\partial}{\partial \cal E}  \right)\frac{\partial}{\partial t} F_{0,ext} \;\; , \label{eq:nuext}
\end{eqnarray}
where $\partial_t F_{0,ext}$ is the rate of change of the fast particle distribution function due to external sources (inclusive of Coulomb collisions). 

Following the same formal steps adopted for the derivation of Eq.~(\ref{eq:deltawknl}), we can obtain the expression of  ${\rm I} \! {\rm Re} \delta\hat{W}_{k, NL}$ at the next order in the asymptotic expansion in $|\gamma/\omega|\approx 1/|\omega\tau_{NL}|$, where $\tau_{NL}$ is the nonlinear time scale:
\begin{eqnarray}
& &\hspace*{-1.2cm} \frac{\partial}{\partial t} {\rm I} \! {\rm Re} \delta \hat{W}_{k,NL}  \simeq - 2 \frac{\pi^2}{B_0^2} m\omega_c^2 \omega \frac{R_0}{r_s^2} \int_0^{r_s} \frac{r^2}{q} dr \int  d {\cal E} d\lambda \sum_{v_\parallel/|v_\parallel| = \pm 1} \frac {\tau_b}{\bar{\omega}_d} \,  \overline{e^{{\rm i} \theta}\omega_d e^{-{\rm i} q(r) \theta}}  \nonumber \\
& & \hspace*{-0.6cm} \times \overline{e^{{\rm i} q(r)\theta}\omega_d e^{-{\rm i} \theta}}  \frac{\partial}{\partial r} \left\{ k_\theta \frac{\partial}{\partial r} \left[ \frac{1}{{\cal E}^{1/2}} \frac{\partial}{\partial \cal E} \left( {\cal E}^{5/2} \overline{e^{{\rm i}(1-q)\theta}\left( 1 - \frac{k_\| v_\parallel}{\omega} \right)} \right. \right. \right. \nonumber \\
& & \hspace*{-0.6cm} \times \left. \overline{e^{{\rm i} q(r)\theta}\omega_d e^{-{\rm i} \theta}} QF_0 \left)   {\rm I} \! {\rm Im} \left( \frac{1}{\bar{\omega}_d - \omega} \right)  r^2 r_s^2 \left| \delta \xi_0\right|^2 \right] \right\} \;\; . \label{eq:redeltawknl}
\end{eqnarray}
The real frequency of the fishbone mode in the nonlinear regime is still given by an equation in the form of
Eq.~(\ref{eq:fishre}): i.e., the mode frequency is expected to {\em chirp} downward as the fast particles relax, according to Eq.~(\ref{eq:ndiff}). More specifically, the nonlinear evolution equation for  the real frequency is
\begin{equation}
\delta \hat {W}_f +  \delta \hat {W}_{f,NL} + {\rm I} \! {\rm Re} \delta \hat {W}_k  + {\rm I} \! {\rm Re} \delta \hat {W}_{k,NL}  = (S/\sqrt{2}) \Lambda^{3/2} \simeq 0 \;\; . \label{eq:fishrenl}
\end{equation}
The effect of external sources (inclusive of Coulomb collisions), can be straightforwardly added into Eq.~(\ref{eq:fishrenl}) via time varying ${\rm I} \! {\rm Re} \delta\hat{W}_{k}$ and $\delta \hat {W}_f$ due to $\partial_t F_{0,ext}$, similarly to Eqs.~(\ref{eq:dbetares}) and~(\ref{eq:nuext}) above. Detailed discussions of these issues will be given elsewhere.

That $\delta \hat{W}_{k,NL}$ is predominantly imaginary, as it emerges from comparisons of Eqs.~(\ref{eq:deltawknl}) and~(\ref{eq:redeltawknl}), suggests that the nonlinear fishbone cycle is essentially determined by the fast particle scattering out of the resonant region. Given the fluctuation level of the mode, the nonlinear time scale, $\tau_{NL}$, as derived from Eq.~(\ref{eq:dbetares}), scales as $(2C)^{1/2} \omega \tau_{NL} \approx \left| \delta \xi_0\right|^{-1}$, consistent with the predator-prey model for the fishbone cycle proposed in \cite{chen84} and in contrast with the time scale $\propto \left| \delta \xi_0\right|^{-2}$ of Eq.~(\ref{eq:redeltawknl}). Meanwhile, the $\propto \left| \delta \xi_0\right|^{-2}$ scaling of characteristic times~\cite{coppi86} is consistent with the time behavior of Eqs.~(\ref{eq:ndiff}) and~(\ref{eq:fishrenl}), describing, respectively, the fast particle relaxation and the time-dependent nonlinear frequency shift of the mode. This picture of the fishbone nonlinear dynamics agrees well with the {\em mode-particle pumping} process, originally proposed in Ref.~\cite{white83}.

Equations~(\ref{eq:fishimnl}) to~(\ref{eq:fishrenl}) describe the bursting fishbone cycle when the dynamics is dominated by coherent nonlinear interactions, typical of the mode-particle pumping process~\cite{white83}, in the presence of quasilinear wave-particle resonances. This formal analysis, thus, is equivalent in spirit to the approach of \cite{breizman98} and the numerical analysis of \cite{candy99}, but has the advantage of treating explicitly the energetic particle nonlinear dynamics. They can be analyzed with different levels of complexity and their detailed analyses will be reported elsewhere, along with comparisons with FTU experimental observations. 
Here, we want to emphasize that Eqs.~(\ref{eq:fishimnl}) and~(\ref{eq:dbetares}) are already a simple yet relevant model which describes the fishbone cycle when the dynamics due to nonlinear frequency shifts is neglected~\cite{chen84,coppi86}. In ~\ref{app:predator}, we show that these equations can be reduced to a predator-prey system, similar to the ad-hoc model introduced in Ref.~\cite{chen84}.
The main results of that analysis are that the nonlinear system is characterized by small oscillation about a fixed point. For increasing LH power input, the system approaches a limit cycle of period $t_{fb}\sim 2\pi/\left(2\Gamma\beta_{h,c}\nu_{ext}\right)^{1/2}$, with $\delta \beta_h/\beta_{h, c}\approx \nu_{ext}^{1/2}/(2\Gamma \beta_{h,c})^{1/2}$ estimating the loss of fast particle in one fishbone burst. Given these results, the present estimate of $t_{fb}$ is consistent with that of Ref.~\cite{chen84}, $t_{fb} \approx (\delta \beta_h/\beta_{h, c}) \nu_{ext}^{-1}$. With the parameters corresponding to the high power phase, $P_{LH}=1.69$ MW, of FTU shot \# 20865, we obtain $t_{fb}\sim 5\div 10$ ms. The good agreement we obtain on the estimate of the fishbone period (no measurements are available of the losses in the perpendicular supra-thermal electron tail) motivates further experimental investigations for more detailed comparisons with theoretical model predictions. 

As in Section~\ref{sec:linear}, we may discuss our conjecture of the relevance of electron fishbone experimental studies for gaining insights into linear and nonlinear burning plasma dynamics. As stated already, the symmetry breaking between fast electron and fast ion bounce averaged dynamics is caused by finite orbit width effects (linear dynamics). In this respect, the typically small dimensionless orbits of fast electrons may generate behaviors analogous to those of well confined fast ions in thermonuclear plasmas. When analyzing nonlinear fast particle behaviors, another source of symmetry breaking between bounce-averaged fast electron and ion dynamics emerges from Eqs.~(\ref{eq:deltahnl}), (\ref{eq:deltawknl}) and~(\ref{eq:redeltawknl}): the term $\propto \overline{v_\parallel \exp {\rm i} (1-q) \theta}$ ($\overline{v_\parallel \exp {\rm i} (1-q) \theta}=0$ for trapped particles), is responsible for the barely circulating fast particle radial transport due to the magnetic component of the fluctuations and clearly depends on the particle mass. For particle distribution functions that are symmetric in $v_\|$, this term is unimportant. Clearly it is not so for the LH driven fast electron distribution, producing a perpendicular fast particle tail, which is moderately slanted toward the counter-current direction. This physics is embedded in the $C$ factor, introduced in Eq.~(\ref{eq:deltawknlsimp}): experimentally varying the power mix of LH and ECRH can, thus, control both the excitation condition of electron fishbones as well as the amount of radial transport due to the magnetic component of the fluctuation. 

\section{Discussions and conclusions}
\label{sec:end}

In this work, we have analyzed the excitation of electron fishbones by both trapped as well as barely circulating supra-thermal electrons, providing a unified explanation of the various experimental observation of these modes. In particular, we have analyzed the peculiarities of electron fishbone excitation on FTU by LH power only, explaining the different roles of trapped and circulating supra-thermal electron tails. The possibility of exciting fishbone modes at frequencies just below the BAE accumulation point by both fast electrons and ions is also discussed and conjectured as interpretation of the experimental evidence of ICRH driven fishbone modes in JET, accumulating at finite frequency above the diamagnetic gap as the mode drive is weakened.

We have derived nonlinear amplitude equations, which describe the nonlinear bursting fishbone cycle due to the mode-particle pumping process in the presence of quasilinear wave-particle resonances when mode-mode couplings are neglected. These equations are qualitatively equivalent to a predator-prey like system, whose predictions are consistent with the corresponding ad-hoc model, originally proposed for explaining the ion fishbone cycle. 

The most interesting feature of electron fishbones is their relevance to burning plasmas. In fact, unlike fast ions in present day experiments, fast electrons are characterized by small orbits, which do not introduce additional complications in the physics due to nonlocal behaviors, similarly to alpha particles in reactor relevant conditions. Meanwhile,  the bounce averaged dynamics of both trapped as well as barely circulating electrons  depends on energy (not mass): thus, their effect on low frequency MHD modes can be used to simulate/analyze the analogous effect of charged fusion products. 
Symmetry breaking between fast electron and ion bounce averaged dynamics is caused by finite orbit width effects (linear dynamics) and by radial transports due to the magnetic component of the fluctuations (nonlinear dynamics of the barely circulating particles). 
In this respect, the combined use of ECRH and LH provide extremely flexible  tools to investigate various nonlinear behaviors, of which  FTU experimental results provide a nice and clear example (see Figure~1). %(see Figure~\ref{fig:figure1}).

\newpage

\noindent{\bf Acknowledgments}

The authors are indebted to useful and stimulating discussions with J.P. Graves, R.J. Hastie, X. Garbet, P. Maget, R.B. White and S.D. Pinches. This work was supported by the Euratom Communities under the contract of Association between EURATOM/ENEA. This work was also supported by the U.S. DOE Contract No. DE-AC02-CHO-3073 and by the Guangbiao Foundation of Zhejiang University. A.V.M. was supported by the Norwegian Research Council under the project No. 171076/V30

\appendix

\section{Generalized inertia and the peculiar roles of trapped and circulating particles}
\label{app:inertia}

Here, we further discuss the peculiar roles of trapped and circulating particles in determining the generalized plasma inertia, continuing the analyses of Section~\ref{sec:inertia}. For this scope, we use the analogy between ZF polarizability and shear Alfv\'{e}n wave inertia enhancement in the banana regime, as in Eq.~(\ref{eq:lambdalow}). Closely following Refs.~\cite{rosenbluth98,hinton99}, we can identify the different roles of trapped and barely circulating particles in determining the $1.6 (R_0/r)^{1/2} q^2$ factor in Eq.~(\ref{eq:lambdalow})~\cite{graves00}. It is easily shown that the inertia enhancement can be written in compact form as:
\begin{equation}
\frac{\omega^2}{\omega_A^2} \left( 1 - \frac{\omega_{*pi}}{\omega} \right) \partial_r^2 \delta \phi \rightarrow \left[  \frac{\omega^2}{\omega_A^2} \left( 1 - \frac{\omega_{*pi}}{\omega} \right) + \Delta I \right] \partial_r^2 \delta \phi \;\; , \label{eq:enhance0}
\end{equation}
where
\begin{eqnarray}
\Delta I \partial_r^2 \delta \phi &=& -  \oint \frac{d\theta}{2\pi} \left\langle \frac{4 \pi }{c^2} e q^2 R_0^2  \omega \omega_d \delta K \right\rangle = \nonumber \\
& & - \frac{4 \pi }{c} {\rm i} q^2 R_0^2 \frac{\omega}{r} \frac{m}{B_0} \int {\cal E} d {\cal E} d\lambda \sum_{v_\parallel/|v_\parallel| = \pm 1} \oint d\theta |v_\|| \frac{\partial}{\partial \theta} \frac{\partial}{\partial r} \delta K \;\; , \label{eq:enhance1}
\end{eqnarray}
where, in the layer, we have used $\omega_d \simeq (v_\| B_0)/(r\omega_c) \partial_\theta (v_\|/B) (-{\rm i} \partial_r)$. Meanwhile, at the leading order for $|\omega/\omega_b|\ll 1$:
\begin{equation}
\sum_{v_\parallel/|v_\parallel| = \pm 1} \oint d\theta |v_\|| \frac{\partial}{\partial \theta} \frac{\partial}{\partial r} \delta K   \simeq  qR_0 \sum_{v_\parallel/|v_\parallel| = \pm 1} \oint d\theta \frac{v_\|}{|v_\||} {\rm i} \omega \frac{\partial}{\partial r} \delta K^{(0)} \;\; . \label{eq:enhance2}
\end{equation}
Here, the lowest order solution $\delta K^{(0)}$ in the $|\omega/\omega_b|$ asymptotic expansion is~\cite{graves00}:
\begin{equation}
\delta K^{(0)}  = \frac{c}{B_0} q \frac{R_0}{r} \tilde{v}_\| \frac{Q F_0}{\omega} \frac{\partial}{\partial r} \delta \phi \;\; , \label{eq:enhance3}
\end{equation}
with $ \tilde{v}_\|$ the fluctuating component of the parallel velocity, defined such $\overline{ \tilde{v}_\|}=0$. Using Eqs.~(\ref{eq:enhance1}) to~(\ref{eq:enhance3}), we readily obtain~\cite{graves00}
\begin{equation}
\Delta I = q^2 \frac{\omega^2}{\omega_A^2} \left( 1 - \frac{\omega_{*pi}}{\omega} \right) \left( \frac{R_0}{r} \right)^{1/2}  f\left(\frac{r}{R_0}\right) \;\; , \label{eq:enhance4}
\end{equation}
where, at the lowest order in $(r/R_0)$,
\begin{eqnarray}
f\left(\frac{r}{R_0}\right) \simeq 1.6 & \simeq & \frac{6\sqrt{2}}{\pi}\int_{\delta}^1 \frac{d\kappa^2}{\kappa^5} \left[ {\rm I} \! {\rm E} \left(\kappa\right) - \frac{\pi^2}{4 {\rm I} \! {\rm K} \left(\kappa\right)} \right] + \frac{3}{8\sqrt{2}} \delta^{1/2} \nonumber \\
& & + \frac{6\sqrt{2}}{\pi}\int_1^\infty \frac{d\kappa^2}{\kappa^6} \left[ (1-\kappa^2) {\rm I} \! {\rm K} \left(1/\kappa\right) + \kappa^2 {\rm I} \! {\rm E} \left(1/\kappa\right) \right] \;\; . \label{eq:trappinertia}
 \end{eqnarray}
Here, $\delta = {\rm O}[(r/R_0)^{1/2}]$; thus, the first two terms on the right hand side (RHS) represent the contribution of barely circulating particles ($\simeq 0.43$), while the last term on the RHS ($\simeq 1.20$) comes from trapped particles. Note that the structure of Eq.~(\ref{eq:trappinertia}) is the same as that involved in the ZF polarizability~\cite{rosenbluth98,hinton99}, as expected.

At low frequency, $|\omega| \laeq \omega_{bi} \approx (r/R_0)^{1/2} \omega_{ti}$, ion Landau damping is strongly decreased due to particle trapping~\cite{bondeson96}; meanwhile, the inertia enhancement due to (well) circulating particles reduces to the $0.5q^2$ factor in Eq.~(\ref{eq:lambdalow}). Note that, as pointed out in Section~\ref{sec:inertia}, the origin of this factor is crucially related to the anisotropic pressure response due to geodesic curvature couplings. In fact, neglecting diamagnetic frequency for simplicity, the usual inertia term $\propto k_r^2 (\omega^2/\omega_A^2) \delta \phi(r)$ is changed into (see Eq.~(\ref{eq:enhance1}))
\begin{equation}
 k_r^2 \frac{\omega^2}{\omega_A^2} \delta \phi(r) + k_r^2 \frac{\omega \omega_{ti}}{\omega_A^2} q^2 \left( \delta \hat{P}_{\| i} + \delta \hat{P}_{\bot i} \right) \;\; , \label{eq:zerofive}
\end{equation}
where $\delta \hat{P}_{\| i}$ and $\delta \hat{P}_{\bot i}$ are the normalized amplitudes of the $\propto \sin \theta$ thermal ion parallel and perpendicular pressure perturbations due to geodesic curvature. For $|\omega| \ll \omega_{ti}$ one easily finds~\cite{zonca96}
\begin{eqnarray}
 \delta \hat{P}_{\| i} & = & \frac{\omega}{\omega_{ti}} \delta \phi(r) \;\; , \nonumber \\
 \delta \hat{P}_{\bot i} & = & - \frac{\omega}{2 \omega_{ti}} \delta \phi(r) \;\; . \label{eq:presslow}
\end{eqnarray}
Thus, the $0.5q^2$ factor is obtained because $\delta \hat{P}_{\bot i}\neq \delta \hat{P}_{\bot i}$, while assuming $\left( \delta \hat{P}_{\| i} + \delta \hat{P}_{\bot i} \right) = 2 \delta \hat{P}_{\| i} = 2 \delta \hat{P}_{i}$ would give the usual $2q^2$ factor~\cite{glasser75}.

\section{A predator-prey like model for the fishbone cycle}
\label{app:predator}

Predator-prey models for the fishbone cycle are known since the original works on the resonant continuum~\cite{chen84} and discrete gap~\cite{coppi86} fishbone modes.
Here, we demonstrate that the nonlinear model equations for the fishbone cycle, Eqs.~(\ref{eq:fishimnl}) to~(\ref{eq:fishrenl}), can be reduced to a structurally stable predator-prey like model with a stable limit-cycle behavior. This gives us a qualitative picture of the fishbone dynamics, which agrees with experimental observations on FTU and with earlier work in Ref.~\cite{chen84}. 
%Thus,  in our opinion, this derivation provides the mathematical and physical foundation of previously proposed phenomenological models~\cite{chen84,coppi86}.
The derivation of a structurally stable model for the fishbone cycle offers the mathematical and physical foundation of previously proposed phenomenological models~\cite{chen84,coppi86}.

For the sake of simplicity, we neglect the dynamics due to time-dependent frequency shifts, contained in Eq.~(\ref{eq:fishrenl}), and we also assume that $\beta_{h, res}$ can be described by a characteristic spatial scale $\Delta \laeq r_s$, so that the first radial derivative $\partial_r \beta_{h, res} \simeq  \mathcal{B} /\Delta$ and the second radial derivative $\partial^2_r \beta_{h, res} \simeq - \mathcal{B} /\Delta^2$, with $\mathcal{B}$ a characteristic value of $\beta _{h, res}$. Note that the second radial derivative is set negative in order to guarantee the stability of the absorbed LH power density profile.
We take for granted that the main contribution to the integral in Eq.~(\ref{eq:fishimnl}) comes from a shell between $r_s - \Delta$ and $r_s$. As an illustration, the profile in Figure~3 %Figure~\ref{fig:figure3}
suggests that  $r_s/a - \Delta/a \approx 0.2$ with a negative second derivative onwards. Integrating in Eq. (35) from $r_s - \Delta$ to $r_s$ we find 
\begin{equation}
\frac{\partial \mathcal{A}}{\partial t} = -2\Gamma (\beta - \mathcal{B}) \mathcal{A} \;\; , \label{eq:b1}
\end{equation}
where we preferred simpler notations $\beta = \beta _{h,c}$ and $\mathcal{A} = |\delta \xi _0| ^2$. Note that the time derivative in Eq.~(\ref{eq:b1}) changes sign when $\mathcal{B}$ crosses the $\beta$ value. If we now turn to Eq.~(\ref{eq:dbetares}) and approximate the radial derivatives of the profile function $\beta _{h, res}$ by their characteristic values through the distance $\Delta$, we get 
\begin{equation}
\frac{\partial}{\partial t} \mathcal{A} \left(\frac{\partial}{\partial t} - \nu _{ext}\right) \mathcal{B} = - \frac{q}{\tau _{NL}} \mathcal{A} ^2 \mathcal{B} \;\; , \label{eq:b2}
\end{equation}  
with $q = 2C\omega^2 \tau_{NL} (r_s / \Delta) ^2$. Differentiating on the left of Eq.~(\ref{eq:b2}) we find
\begin{equation}
\frac{\partial \mathcal{A}}{\partial t} \left(\frac{\partial}{\partial t} - \nu _{ext}\right) \mathcal{B} + \mathcal{A}\frac{\partial}{\partial t} \left(\frac{\partial}{\partial t} - \nu _{ext}\right) \mathcal{B} = - \frac{q}{\tau _{NL}} \mathcal{A} ^2 \mathcal{B} \;\; . \label{eq:b3}
\end{equation} 
We are interested in the asymptotic dynamics when the time $t\rightarrow +\infty$ and we bypass the intermediate transitional-type patterns of behavior when the system basically evolves from a linear starting regime into a strongly nonlinear stage when the coupling between the dynamical parameters comes into play.
Keeping first time-derivatives and suppressing higher-order differential terms we rewrite Eq.~(\ref{eq:b3}) as
\begin{equation}
\frac{\partial \mathcal{A}}{\partial t} \left(\frac{1}{\tau_{NL}} - \nu_{ext}\right) \mathcal{B} + \frac{\mathcal{A}}{\tau_{NL}} \left(\frac{\partial}{\partial t} - \nu_{ext}\right) \mathcal{B} = - \frac{q}{\tau_{NL}} \mathcal{A} ^2 \mathcal{B} \;\; , \label{eq:b4}
\end{equation} 
where $1 / \tau _{NL}$ stands for $\partial / \partial t$ where appropriate to constitute the right ordering. In writing Eq.~(\ref{eq:b4}), we considered that the dynamical time scale is of the same order of the nonlinear time $\tau_{NL}$ so that the nonlinearities are essentially present through the dynamics. Note that Eq.~(\ref{eq:b4}) does not admit the linear limit as particular case. We also took into account that the time derivative $\partial\mathcal{A} / \partial t$ is such as to satisfy the dynamic Eq.~(\ref{eq:b1}). Substituting $\partial\mathcal{A} / \partial t$ from Eq.~(\ref{eq:b1}), after simple algebra one obtains  
\begin{equation}
\frac{\partial \mathcal{B}}{\partial t} = \nu\mathcal{B} - \Theta \mathcal{B}^2 - q\mathcal{A}\mathcal{B} \;\; , \label{eq:b5}
\end{equation}  
with $\nu = \nu_{ext} + \beta\Theta$ and $\Theta = 2\Gamma (1 - \nu_{ext} \tau_{NL})$. Equations~(\ref{eq:b1}) and~(\ref{eq:b5}) form a predator-prey system of equations. If we change the notations in Eqs.~(\ref{eq:b1}) and~(\ref{eq:b5}) such that $\mathcal{A}$ is $x$ and $\mathcal{B}$ is $y$ and introduce the parameters $\mu = 2\beta\Gamma$ and $k = 2\Gamma$, we can represent our predator-prey model in the canonical form 
\begin{equation}
\dot{x} = -\mu x + kxy \;\; , \label{eq:b6}
\end{equation}
\begin{equation}
\dot{y} = \nu y -\Theta y^2 - qxy \;\; , \label{eq:b7}
\end{equation}
where the dot denotes time differentiation. Equations~(\ref{eq:b6}) and~(\ref{eq:b7}) define a dynamical system with an unstable hyperbolic point at the origin and an equilibrium (i.e., fixed point) at $x_0 = \nu / q - \Theta \mu / qk$ and $y_0 = \mu / k$. 
The unstable hyperbolic point at the origin, $(x,y)=(0,0)$, is a signature of the linear instability of the system. Meanwhile, the fixed point at $(x,y)=(x_0,y_0)$ dictates the nonlinear behaviors.
The term with $y^2$ in Eq.~(\ref{eq:b7}) is important as it guarantees the structural stability of the model, in the topological sense. If $\Theta > 0$, the system shows a stable limit cycle behavior, in agreement with the bursting mode signatures in FTU, as discussed below. 
A transition to the limit-cycle dynamics corresponds to a nonlinear time $\tau_{NL}$, which is comparable to or shorter than $\nu_{ext}^{-1}$. Physically, this means that the system accommodates the external changes in the profile function and, thus, kind of {\em digests} the external power density input.
If one wishes to obtain a marginal estimate, then the procedure is to let $\tau_{NL}$ be of the order of $\nu_{ext}^{-1}$ and neglect the term with $\Theta$ in Eq.~(\ref{eq:b7}), yielding    
\begin{equation}
\dot{x} = -\mu x + kxy \;\; , \label{eq:b8}
\end{equation}
\begin{equation}
\dot{y} = \nu y - qxy \;\; , \label{eq:b9}
\end{equation}
with the fixed point at $x_0 = \nu / q$ and $y_0 = \mu / k$. A perturbation analysis of the reduced Eqs.~(\ref{eq:b8}) and~(\ref{eq:b9}) shows that the dynamics are periodic, with frequency $\Omega = (\mu\nu)^{1/2}=(2\Gamma\beta_{h,c}\nu_{ext})^{1/2}$. The trajectories of the system in the $(x,y)$ phase space are defined through~\cite{white89} 
\begin{equation}
x^{1/\mu}y^{1 / \nu} \exp \left(-\frac{qx + ky}{\mu\nu}\right) = {\rm const} \label{eq:phase}
\end{equation}
and are plotted numerically in Figures~6 to 8 %Figures~\ref{fig:figure6} to~\ref{fig:figure8}
for different values of the ratio $\alpha=\mu / \nu=(2\Gamma\beta_{h,c}/\nu_{ext})$ and with normalized axis scales $(x/x_0,y/y_0)$: in this way the fixed point is always (1,1). Given $\delta x$ and $\delta y$ the excursions about the fixed point with frequency $\Omega$, we typically have $\delta y/y_0 \sim \nu_{ext}\Omega ^{-1}$ for $\delta x/x_0 \sim 1$. This means that the characteristic excursion of $\beta_h$ about the fixed-point value is $\delta \beta_h / \beta _{h, c} \sim (\nu_{ext} / 2\Gamma\beta _{h, c}) ^{1/2}$. With these estimates, the period of fishbone burst is $t_{fb} \sim 2\pi / \Omega \sim 2\pi (\delta \beta _h / \beta _{h, c}) \nu_{ext} ^{-1}$, in agreement with the estimate given in Ref.~\cite{chen84}. Note that the wider the oscillation amplitude around the fixed point is, the more important the non-harmonic behavior becomes in the periodic motion of the system, as it is clearly visible in Figures~6 to 8, %Figures~\ref{fig:figure6} to~\ref{fig:figure8},
consistently with the electron fishbone burst signature of Figure~1. %Figures~\ref{fig:figure1}.
The nonlinear excursions of the system about the fixed point have an amplitude which is dictated by the external power density input into the wave-particle resonance region, i.e., {\it ceteris paribus}, by $\nu_{ext}$, which is experimentally controlled via the additional power level. 

 \begin{figure}
 \begin{center}
 \noindent\epsfxsize=0.75\linewidth\epsfbox{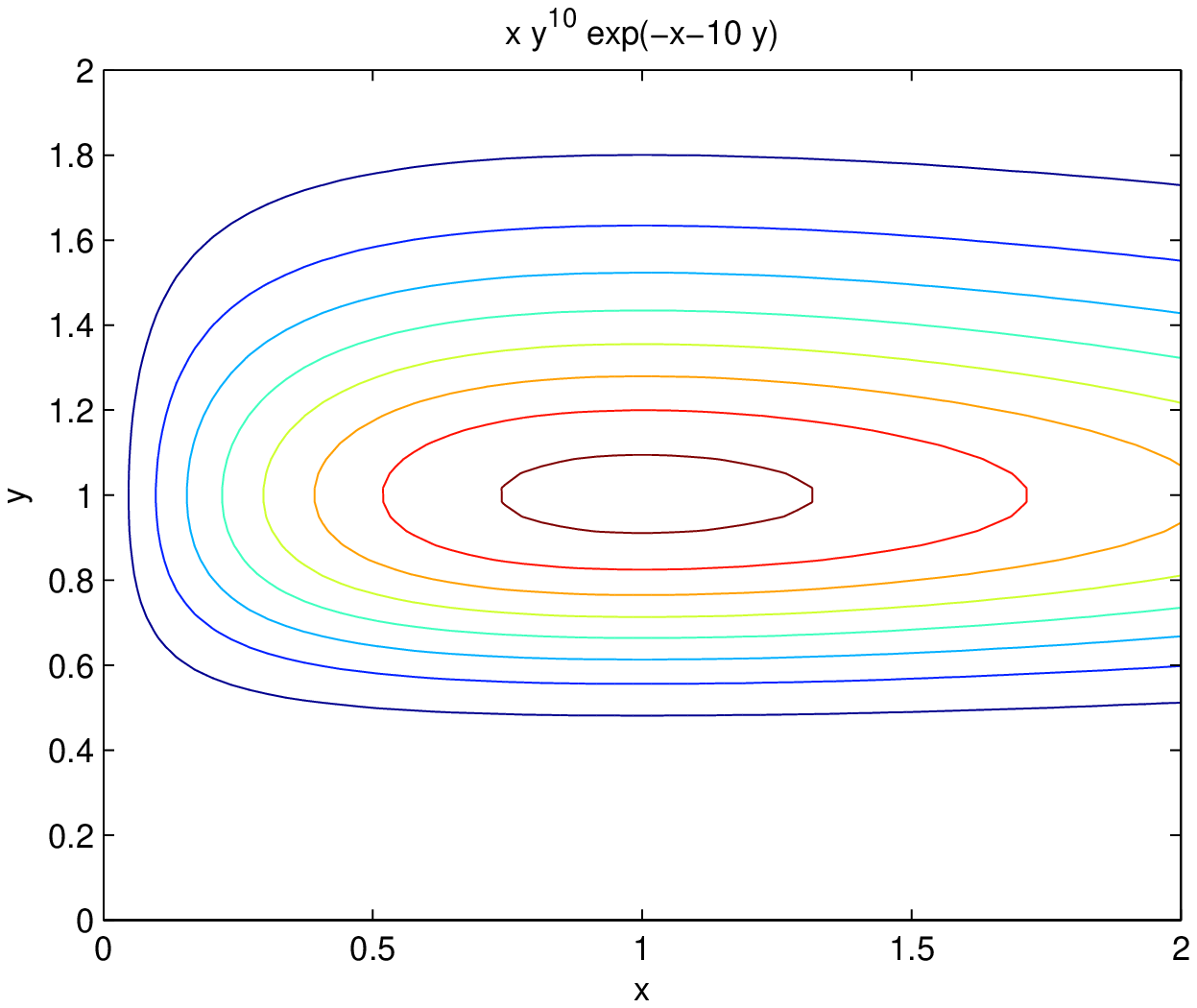}
 \end{center}
 \label{fig:figure6}
 \caption{Contour plot of Eq.~(\ref{eq:phase}) trajectories in the $(x/x_0,y/y_0)$ phase space. Here, $\alpha=2\Gamma \beta_{h,c}/\nu_{ext} = 10$.}  
 \end{figure}
 
 \begin{figure}
 \begin{center}
 \noindent\epsfxsize=0.75\linewidth\epsfbox{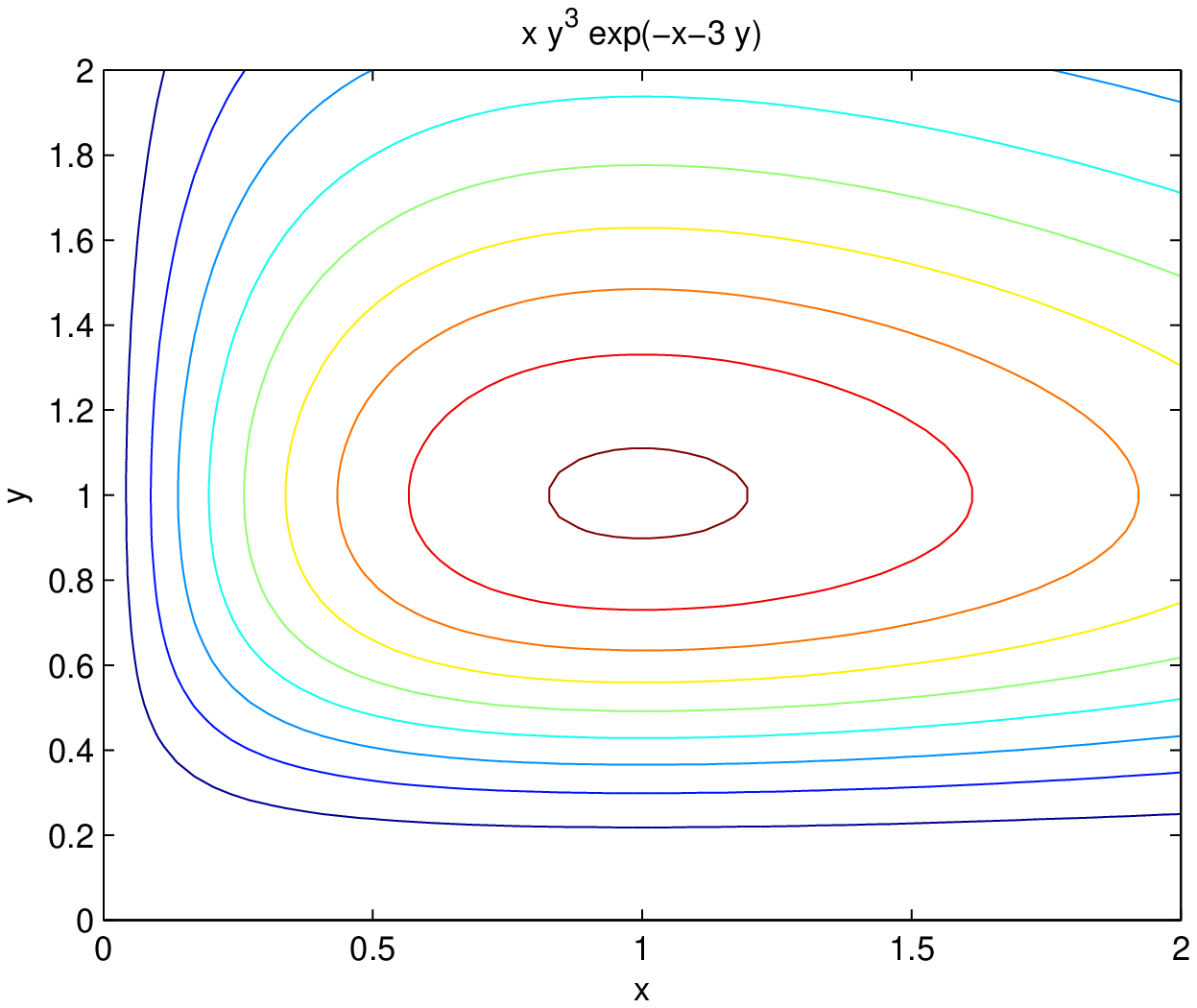}
 \end{center}
 \label{fig:figure7}
 \caption{Contour plot of Eq.~(\ref{eq:phase}) trajectories in the $(x/x_0,y/y_0)$ phase space. Here, $\alpha=2\Gamma \beta_{h,c}/\nu_{ext} = 3$.}  
 \end{figure}
 
 \begin{figure}
 \begin{center}
 \noindent\epsfxsize=0.75\linewidth\epsfbox{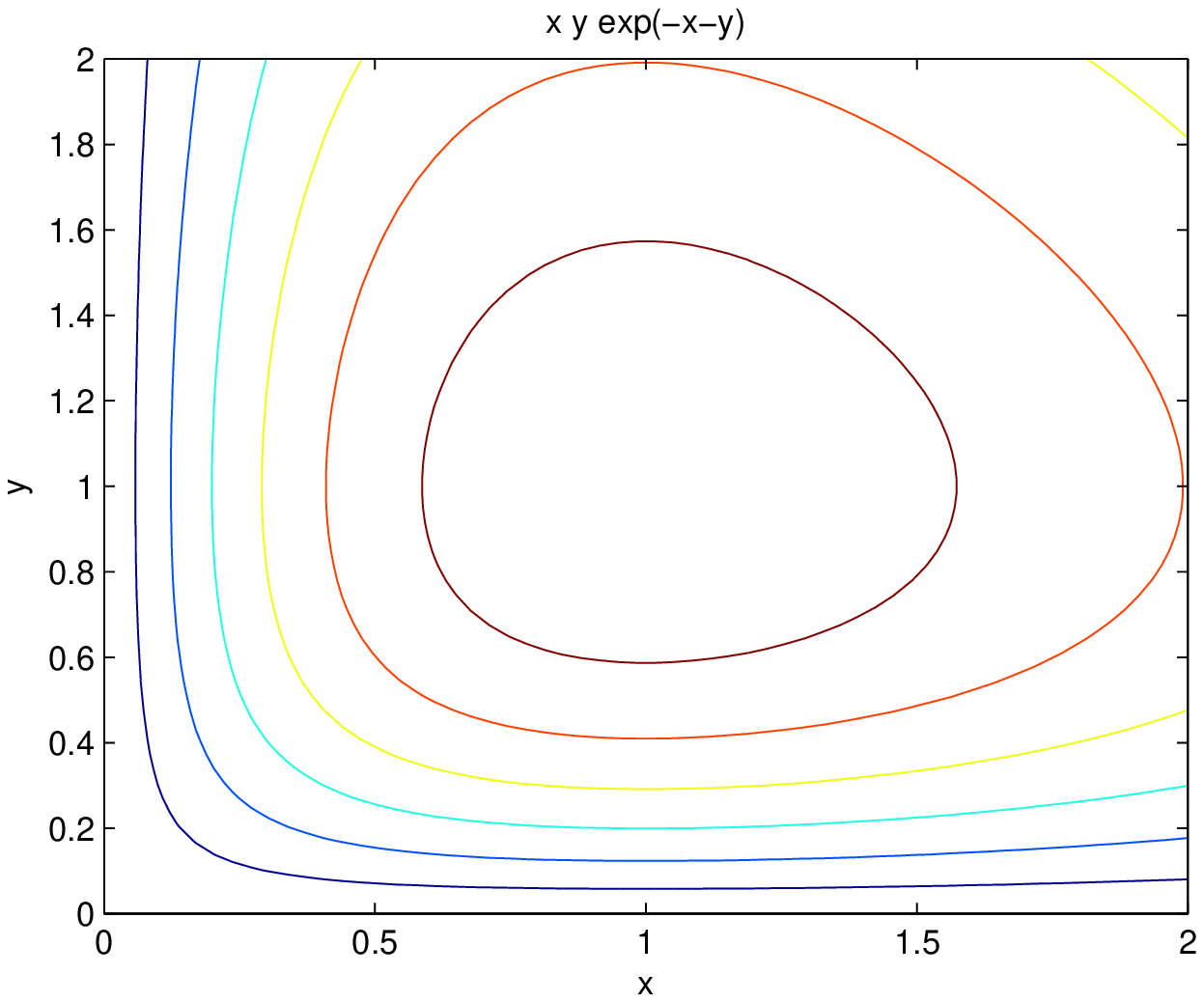}
 \end{center}
 \label{fig:figure8}
 \caption{Contour plot of Eq.~(\ref{eq:phase}) trajectories in the $(x/x_0,y/y_0)$ phase space. Here, $\alpha=2\Gamma \beta_{h,c}/\nu_{ext} = 1$.}  
 \end{figure}

\end{document}